\documentclass[10pt,final, twocolumn]{IEEEtran}
\IEEEoverridecommandlockouts
\def\BibTeX{{\rm B\kern-.05em{\sc i\kern-.025em b}\kern-.08em
    T\kern-.1667em\lower.7ex\hbox{E}\kern-.125emX}}

\usepackage{amsmath,algorithm, algorithmic}
\usepackage{cite}
\usepackage{epsf,graphics,graphicx}
\usepackage{comment}
\usepackage{bm}
\usepackage{amssymb,amsthm}
\usepackage{amsmath}
\usepackage[top=.75in, left= 0.625in, right = 0.625in, bottom =1in]{geometry}

\newtheorem{assumption}{Assumption}

\newtheorem{identity}{Identity}

\newtheorem{Proposition}{Proposition}
\newtheorem{remark}{Remark}
\newtheorem*{remark*}{Remark}

\usepackage{comment}
\usepackage[textwidth=30mm]{todonotes}
\usepackage{soul}

\setstcolor{magenta}
\sethlcolor{lightgray}

\def\vec#1{{\bf #1}}
\def\Tr#1{\mathrm{Tr}\left\{ #1 \right\} }
\def\outer{\otimes}
\def\bm#1{{\mbox{\boldmath{$#1$}}}}

\DeclareMathOperator{\ex}{\mathbb{E}}

\newcommand{\ZZ}{\mathbb{Z}}

\newcommand{\CC}{\mathbb{C}}

\def\bPhi{\mbox{\boldmath$\Phi$}}
\def\bSigma{\mbox{\boldmath$\Sigma$}}

\def\bA{{\bf A}}
\def\bB{{\bf B}}
\def\bC{{\bf C}}

\def\bG{{\bf G}}

\def\bI{{\bf I}}
\def\bQ{{\bf Q}}
\def\bR{{\bf R}}
\def\bS{{\bf S}}
\def\bT{{\bf T}}

\def\bV{{\bf V}}
\def\bW{{\bf W}}

\def\bY{{\bf Y}}
\def\bZ{{\bf Z}}

\def\nt{{N_t}}
\def\nr{{N_r}}
\def\ns{{N_s}}
\def\nd{{n_d}}

\def\bRcal{\mathbfcal{R}}
\def\bTcal{\mathbfcal{T}}

\DeclareMathAlphabet\mathbfcal{OMS}{cmsy}{b}{n}

\def\be{{\bf e}}
\def\bk{{\bf k}}

\def\bn{{\bf n}}
\def\bm{{\bf m}}

\def\bp{{\bf p}}
\def\bq{{\bf q}}

\def\bs{{\bf s}}

\def\bu{{\bf u}}
\def\bv{{\bf v}}

\def\bx{{\bf x}}
\def\by{{\bf y}}
\def\bz{{\bf z}}

\begin{document}
\bstctlcite{BSTcontrol}

\title{Reconfigurable Intelligent Surfaces and\\ Capacity Optimization: A Large System Analysis}

\author{Aris~L.~Moustakas,~\IEEEmembership{Senior~Member,~IEEE,} George~C.~Alexandropoulos,~\IEEEmembership{Senior~Member,~IEEE,}\\ and M\'{e}rouane Debbah,~\IEEEmembership{Fellow,~IEEE}
\thanks{Part of this paper was presented in IEEE GLOBECOM, Madrid, Spain, 7--11 December 2021 \cite{Moustakas_RIS}. This work has been supported by the EU H2020 RISE-6G project under grant number 101017011.}
\thanks{A.~L.~Moustakas is with the Department of Physics, National and Kapodistrian University of Athens, 15784 Athens, Greece (e-mail: arislm@phys.uoa.gr).}
\thanks{G.~C.~Alexandropoulos is with the Department of Informatics and Telecommunications, National and Kapodistrian University of Athens, 15784 Athens, Greece and also with the Technology Innovation Institute, 9639 Masdar City, Abu Dhabi, United Arab Emirates
(e-mail: alexandg@di.uoa.gr).}
\thanks{M. Debbah is with both the Technology Innovation Institute and the Mohamed Bin Zayed University of Artificial Intelligence, 9639 Masdar City, Abu Dhabi, United Arab Emirates (email: merouane.debbah@tii.ae).}
}

\maketitle

\begin{abstract}
Reconfigurable Intelligent Surfaces (RISs), comprising large numbers of low-cost and almost passive metamaterials with tunable reflection properties, have been recently proposed as an enabling technology for programmable wireless propagation environments. In this paper, we present asymptotic closed-form expressions for the mean and variance of the mutual information metric for a multi-antenna transmitter-receiver pair in the presence of multiple RISs, using methods from statistical physics. While nominally valid in the large system limit, we show that the derived Gaussian approximation for the mutual information can be quite accurate, even for modest-sized antenna arrays and metasurfaces. The above results are particularly useful when fast-fading conditions are present, which renders instantaneous channel estimation extremely challenging. We find that, when the channel close to an RIS is correlated, for instance due to small angle spread, which is reasonable for wireless systems with increasing carrier frequencies, the communication link benefits significantly from statistical RIS phase optimization, resulting in gains that are surprisingly higher than the nearly uncorrelated case. Using our novel asymptotic properties of the correlation matrices of the impinging and outgoing signals at the RISs, we can optimize the metasurfaces without brute-force numerical optimization. Furthermore, when the desired reflection from any of the RISs departs significantly from geometrical optics, the metasurfaces can be optimized to provide robust communication links, without significant need for their optimal placement.
\end{abstract}

\begin{IEEEkeywords}
Reconfigurable intelligent surface, multipath, beamforming, capacity, MIMO, random matrix theory, replicas.
\end{IEEEkeywords}


\section{Introduction}
Future wireless networks are expected to transform to a unified communication, sensing, and computing platform with embedded intelligence, enabling sixth Generation (6G) applications and service requirements \cite{Akyildiz_6G_2020, Samsung}. To accomplish this overarching goal, advances at various aspects of the network design are necessary, including wideband front-ends and smart wireless connectivity schemes \cite{Samsung}. Reconfigurable Intelligent Surfaces (RISs) \cite{huang2019holographic} constitute a key wireless hardware technology for the recently conceived concept of ElectroMagnetic (EM) wave propagation control \cite{liaskos2018new,huang2019reconfigurable,di2019smart,RIS_Scattering,WavePropTCCN,risTUTORIAL2020}, which is envisioned to offer artificial manipulation of the wireless environment. This low-cost and almost energy-neutral technology enables coating the various obstacles and objects of the environment with ultra-thin metasurfaces, thus, transforming them into network entities with dynamically reconfigurable properties that can facilitate wireless communications, localization, and sensing \cite{RISE6G_COMMAG,RIS_Localization,zhang2022towards}. This technological potential gave rise recently to the ``smart wireless environment as a service'' paradigm \cite{rise6g}, which introduces the ``Layer $0$'' being capable to affect the radio wave propagation environment on top of
which data and control signals are transmitted via physical ``Layer $1$."

Over the last few years, metamaterials have emerged as a powerful technology with a broad range of applications, including wireless communications \cite{WavePropTCCN}. They constitute artificial elements with physical properties that can be engineered to exhibit various desired characteristics (e.g., reflective beamforming, signal absorption, Doppler cloak, and bandwidth narrowing) \cite{smith2,Science_2011}. When deployed in planar structures (a.k.a. metasurfaces), their effective parameters
can be tailored to realize desired reflections, and consequently over-the-air analog processing \cite{RIS_Scattering}, of their impinging EM waves \cite{smith3}. RISs are essentially surfaces comprising many small reflecting meta-elements, which may be tuned independently to manipulate the metasurface's reflection properties. This creates an opportunity to jointly optimize network-controlled metasurfaces with the channel degrees of freedom, through a recently proposed ``holographic'' paradigm of channel description \cite{Holographic_Pizzo, Holographic_WeiLi}. 

Leveraging the aforementioned RIS perspectives, metasurfaces have been recently considered in the wireless communications field for various applications, including signal coverage extension \cite{bjornson_intelligent_2020, yildirim_hybrid_2021}, 
accurate localization boosting \cite{abu-shaban_near-field_2020,locrxris,zerobs}, 
and enabling physical-layer security \cite{Cui_2019,PLS_Kostas}.  
A passive beamformer that can achieve the asymptotic optimal performance by controlling the properties of the incoming EM wave at the RIS was designed in \cite{Jung_2021a}, considering a limited RIS control link and practical reflection coefficients. The fundamental capacity limits of RIS-assisted multi-user wireless communication systems were investigated in \cite{Mu_2021b}. The authors in \cite{You_2021} considered the application of RISs to assist the uplink transmission from multiple users to a multi-antenna base station, and devised an optimization framework for jointly designing the transmit covariance matrices and the RIS phase profiles, for the case where only partial channel state information is available. In \cite{Xu_2021b}, active RISs were considered, which can adapt the phase and amplify the magnitude of the reflected incident signal simultaneously with the support of an additional power source. Hybrid RISs, capable of simultaneously controlling the reflection of a portion of the impinging EM wave, while performing sensing at the remaining portion of it, were proposed in \cite{HRIS_Mag,HRIS_Nature}, and were very recently \cite{HRIS_CE} deployed for facilitating channel estimation in RIS-empowered multi-user uplink communications. In \cite{Yang_2021c}, the problem of energy efficiency optimization for a wireless communication system assisted including distributed RISs was investigated. RIS-empowered device-to-device communications underlaying a cellular network were considered in \cite{Yang_2021d}, in which an RIS was employed to enhance the desired signals and suppress interference between paired device-to-device and cellular links. Multi-RIS smart wireless environments were recently considered in \cite{Samarakoon_2020,pervasive_DRL_RIS}, whose orchestration was based on supervised and unsupervised machine learning tools, leveraging position information for the receiving user \cite{Samarakoon_2020} and instantaneous multi-user channel knowledge availability \cite{pervasive_DRL_RIS}.

Apart from the aforepresented representative algorithmic approaches for the phase configurations of RISs, analyzing the performance gains that metasurfaces can offer in wireless communications has lately attracted research attention, as a means to unveil the true potential of multi-RIS-empowered smart wireless environments for 6G networks, and accordingly, trigger efficient optimization approaches relying on practically acquirable channel state information \cite{Tsinghua_RIS_Tutorial}. An asymptotic analysis of the uplink data rate in a multi-user setup with a single RIS was carried out in \cite{Jung2020}, considering spatially correlated Rician fading channels subject to estimation errors as well as hardware impairments for the metasurface. For the same system model, but for the downlink direction, \cite{Nadeem2020} studied the optimum linear precoding matrix that maximizes the minimum signal-to-interference-plus-noise ratio, and presented deterministic approximations for its parameters. Considering Nakagami-$m$ fading conditions and various performance metrics, analytic approximations for the single-RIS reflecting elements were presented in \cite{Selimis2021}. An asymptotic analysis and approximations for the interference-to-noise ratio considering uncorrelated Rayleigh fading channels were derived in \cite{INR_analysis}, while \cite{analysis_intertwinement} presented upper and lower bounds for the outage probability and ergodic capacity, considering the intertwinement model of \cite{Abeywickrama_2020} between the amplitude and phase RIS response. However, all previously described studies focus on scenarios with a single RIS, consider availability of the instantaneous channels, and overlook the role of the channel conditions near the RIS. 


\subsection{Contributions}
In this paper, we analyze the statistics of the Mutual Information (MI), which serves as a performance metric that unveils the potential gains from the adoption of multiple RISs for assisting a multi-antenna Transmitter (TX) and Receiver (RX) pair. More specifically, this paper's contributions are summarized as follows:
\begin{itemize}
\item
We use tools from random matrix theory and statistical physics to derive analytic expressions for the ergodic MI and its variance, nominally valid in the limit of large antenna numbers and large numbers for the reflecting elements of the RISs. 
\item 
By showing that the distribution of the MI metric converges weakly to the standard normal, we present a sufficiently accurate approximation for the outage MI, which is a relevant communication performance metric for  block-fading channels.
\item
We obtain the asymptotic properties of the correlation matrices corresponding to the incoming and outgoing EM waves to and from the RISs, as well as their relation to their two-dimensional Fourier transform. 
\item 
We optimize the ergodic MI performance subject to knowledge of the statistical properties of the channel, which is a more realistic optimization strategy for the case of multiple RISs, due to their size and the usually present channel fluctuations. In addition, using the knowledge of the asymptotic form of the eigenvectors of the channel correlation matrices, essentially the 
degrees of freedom of the RIS-parametrized channel, we are able to directly optimize the multiple RISs in closed form.

\item We compare our analytic results with numerical optimization algorithms and Monte Carlo simulations, exhibiting exceptional agreement, despite the relatively small size of the TX and RX antenna arrays we use.
\end{itemize}

\subsection{Outline}
In the next Section~\ref{sec:MIMO channel model}, we describe the considered system and channel models. In Section \ref{sec:MI_Analysis}, we provide analytic results for the MI statistics and the asymptotic properties of the correlation matrices. Section~\ref{sec:MI_Optimization} deals with the optimization of the MI with respect to the phase configurations of the multiple RISs and their locations. In Section~\ref{sec:Numerical_Results}, we evaluate our analytic results in various cases comparing them with equivalent from numerical optimization. Finally, in Section~\ref{sec:conclusion}, we present the paper's concluding remarks, while Appendix~\ref{app:proof_ergMI} sketches the proof of our asymptotic results using statistical physics tools.
\subsection{Notations} 
Throughout the paper, we use bold-faced upper-case letters for matrices, e.g., $\vec X$ with its $(i,j)$-element denoted by $[\vec X]_{i,j}$, and bold-faced lower-case letters for column vectors, e.g., $\vec x$ with its $i$-element represented by $[\vec x]_i$. The superscripts $T$ and $\dagger$ indicate
transpose and Hermitian conjugate operations, $\Tr{\, \cdot\,  }$ and $\left\Vert\cdot\right\Vert_F$ denote the trace and Frobenius norm of a matrix, respectively, and $\vec I_n$ ($n\geq2$) represents the $n$-dimensional identity matrix. The superscripts/subscripts $t$ and $r$ are used for quantities (e.g., channel matrices) referring to the Transmitter (TX) and Receiver (RX), respectively. $\mathbf{x}\sim{\cal CN}(\mathbf{0}_n,\mathbf{I}_n)$ denotes a $n$-element complex and circularly symmetric Gaussian vector with zero-mean elements and covariance matrix $\mathbf{I}_n$, while $\ex[\,\cdot\,]$ is the expectation operator. 

\section{System and Channel models}\label{sec:MIMO channel model}
Consider a wireless communication system between a TX equipped with $\nt$ antenna elements and an RX with an $\nr$-antenna array over a fading channel comprising their direct link as well as channel components resulting from reflections from $K$ identical RISs, each consisting of $\ns$ tunable reflecting elements \cite{Tsinghua_RIS_Tutorial}. We assume that all channels are known to the RX via an adequate estimation approach (see \cite{hardware2020icassp,Swindlehurst_CE,HRIS_CE} for different RIS hardware capabilities), but not to the TX. 
The baseband representation of the $\nr$-dimensional received signal vector at RX can be expressed as follows:
\begin{equation}\label{eq:basic_channel_eq}
  \by \triangleq \sqrt{\rho}\bG_{\rm tot}  \bx + \bz,
\end{equation}
where $\bx$ is the $\nt$-dimensional signal vector with covariance matrix $\bQ\triangleq \ex[\bx\bx^\dagger]$ normalized such that $\Tr{\bQ}=\nt$, and $\bz\sim{\cal CN}(\mathbf{0}_\nr,\mathbf{I}_\nr)$ is the noise vector.
The $\nr\times\nt$ channel gain matrix $\bG_{\rm tot}$ in our system model  \eqref{eq:basic_channel_eq} can be written as\footnote{The cascaded end-to-end channel $\bG_{r,k}\bPhi_k\bG_{t,k}$ included in \eqref{eq:Gtot}, which first appeared in \cite{huang2019reconfigurable}, models each TX-RX link enabled by each $k$-th RIS, and has been widely considered up to date for analyzing and optimizing RIS-empowered wireless communications \cite{risTUTORIAL2020}. As discussed in \cite{DRL_automonous_RIS}, it can incorporate mutual coupling among the front-end radiating elements at any of the communicating nodes \cite{alexandg_ESPARs,Gradoni2020}, and exhibits certain connections with recently presented physics-inspired channel models \cite{PhysFad}.}
\begin{figure}[!t]
	\centering
	\includegraphics[width=\columnwidth]{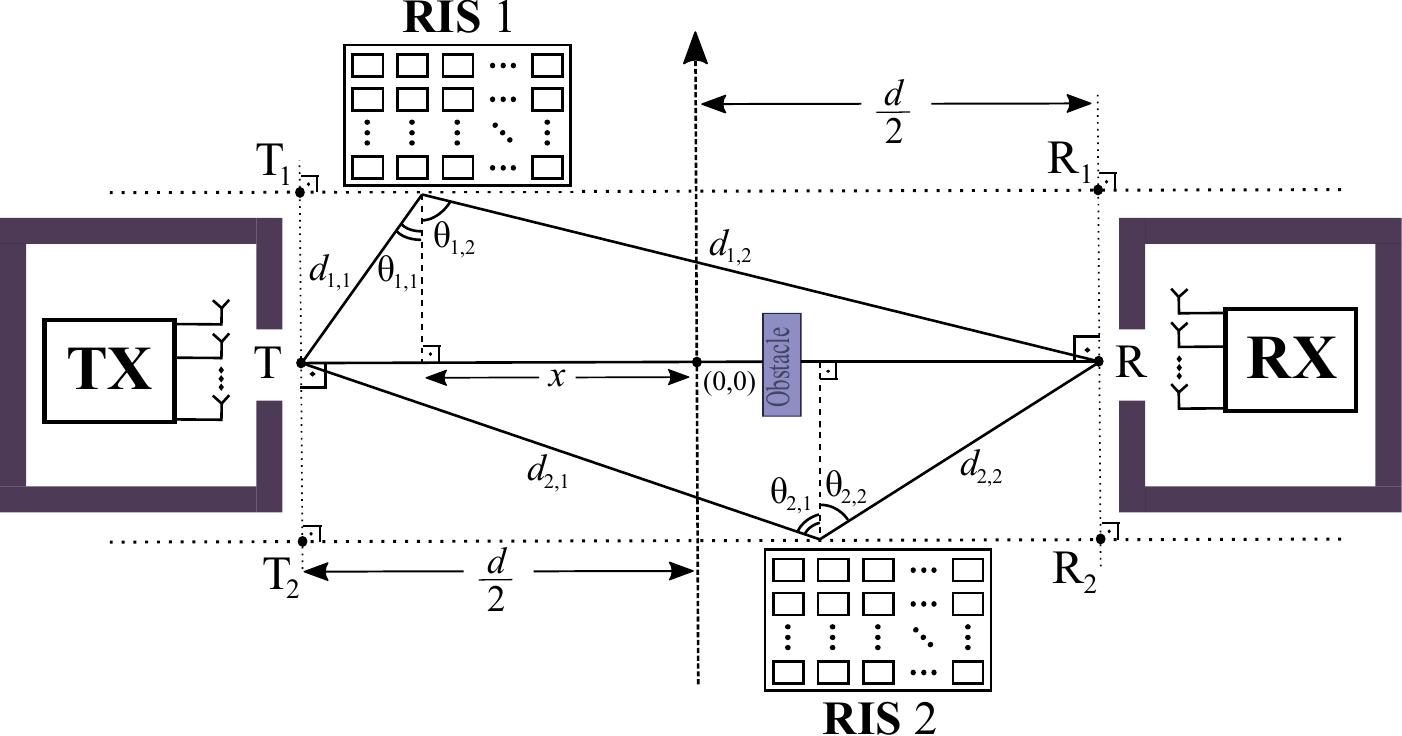}
	  \caption{An example wireless communication system between an $\nt$-antenna TX and an $\nr$-antenna RX empowered by $K=2$ identical and $\ns$-element RISs. The EM wave propagation environment may include obstacles (dark and light grey rectangulars) contributing to local scattering at the TX and RX, and obstructing the direct TX-RX link (i.e., the channel that does not include reflections from the RISs). Notations for the inter-node Euclidean distances as well as the angles of the incoming and outgoing EM waves at an arbitrary reference element at each RIS are included.}
		\label{fig:system_model}
\end{figure}
\begin{align}\label{eq:Gtot}
\bG_{\rm tot} \triangleq \bG_d + \sum_{k=1}^K\sqrt{\gamma_k}\bG_{r,k}\bPhi_k\bG_{t,k},
\end{align}
where $\bG_{r,k}$ and $\bG_{t,k}$, with $k=1,2,\ldots,K$, represent the $\nr\times \ns$ and $\ns\times\nt$ channel matrices from the $k$-th RIS to RX and from the TX to the $k$-th RIS, respectively, while $\bG_d$ denotes the direct $\nr\times\nt$ channel matrix between the RX and TX, which does not include any reflection from any of the RISs. In addition, $\bPhi_k$ is the $\ns$-dimensional diagonal square matrix containing the tunable reflection coefficients at the $k$-th RIS in the main diagonal. The parameter $\rho$ in \eqref{eq:basic_channel_eq} represents the Signal-to-Noise Ratio (SNR) of the total link, while $\gamma_k$ is a measure of the relative additional SNR from each $k$-th RIS. An example of a communication system with $K=2$ RISs appears in Fig$.$~\ref{fig:system_model}, in which each $\gamma_k$ for $k=1$ and $2$ can be expressed as follows:
\begin{align}\label{eq:gamma}
\gamma_k = \left(\frac{\frac{d^2}{4}+h^2}{d_{k,1}d_{k,2}}\right)^{b}.
\end{align}
In this expression, we have assumed, without loss of generality, the same pathloss (far-field) exponent $b$ for all involved links, and $d_{k,\ell}$ with $\ell=1$ and $2$ represents the Euclidean distances between the TX and the $k$-th RIS (first hop) and between the $k$-th RIS and the RX (second hop), respectively. Finally, for simplicity, we have normalized the value of $\gamma_k$ to its value at location $(d/2,h)$ for an arbitrary value of $h$. In our channel model in \eqref{eq:Gtot}, we have made the common reasonable assumption that consecutive reflections from the same or multiple RISs are negligible, due to the large multiplicative pathloss.

The $n$-th reflection coefficient, with $n=1,2,\ldots,\ns$, of the $k$-th RIS is modeled as $[\bPhi_k]_{n,n}\triangleq e^{i\phi_{k,n}}$ \cite{risTUTORIAL2020}; extensions to practical phase shift models \cite{Abeywickrama_2020} and measured element responses \cite{Rahal_Measured_Responses} are left for future work. We also assume that all involved channel matrices in $\bG_{\rm tot}$ are complex Gaussian having the following Kronecker-product-form covariances  $\forall i,j=1,2,\ldots,\nr$, $\forall m,n=1,2,\ldots,\nt$, and $\forall a,b=1,2,\ldots,\ns$:
\begin{subequations}\label{eq:all_correlations}
\begin{align}
\label{eq:Gd_cov}
\ex\left[[\bG_d]_{i,m}[\bG_d]_{j,n}^*\right]&=\frac{1}{\nt}[\bR_d]_{i,j} [\bT_d]_{m,n},
\\ \label{eq:Grk_cov}
\ex\left[[\bG_{r,k}]_{i,a}[\bG_{r,k}]_{j,b}^*\right]&=\frac{1}{\nt}[\bR_k]_{i,j} [\bS_{r,k}]_{a,b},
\\ \label{eq:Gtk_cov}
\ex\left[[\bG_{t,k}]_{a,m}[\bG_{t,k}]_{b,n}^*\right]&=\frac{1}{\nt}[\bS_{t,k}]_{a,b} [\bT_k]_{m,n}.
\end{align}
\end{subequations}
In the above expressions, the received correlation matrices, namely $\bR_k$ of dimension $\nr\times \nr$, $\bS_{r,k}$ of dimension $\ns\times \ns$, and $\bR_d$ of dimension $\nr\times \nr$, as well as the transmit correlation matrices, namely $\bT_k$ of dimension $\nt\times \nt$, $\bS_{t,k}$ of dimension $\ns\times \ns$, and $\bT_d$ of dimension $\nt\times \nt$ are all non-negative definite having the following fixed traces: $\Tr{\bT_k}=\Tr{\bT_d}=\nt$, $\Tr{\bR_k}=\Tr{\bR_k}=\nr$, and $\Tr{\bS_{r,k}}=\Tr{\bS_{t,k}}=\ns$. Note that $\bS_{r,k}$ models the correlation of the incoming EM waves at the elements of each $k$-th RIS, while $\bS_{t,k}$ models the correlation at those elements for the outgoing (reflected) EM waves. For simplicity, we will not consider the polarization properties of the EM waves, treating them here only as scalars. The extension with polarization will be treated in a future work. 

In the considered case, all above correlation matrices may be expressed in terms of weight functions $w(\bk)$ of the incoming or outgoing waves \cite{Moustakas2000_BLAST1_new}, where $\bk$ is the corresponding $3$-dimensional wave vector with magnitude $|\bk|=k_0\triangleq\frac{2\pi}{\lambda}$, where $\lambda$ represents the wavelength. Following the latter notation, each $(a,b)$-element of the matrix $\bS_{r,k}$ $\forall k$ can be respectively obtained as:
\begin{equation}\label{eq:corr_mat_w(k)_def}
    \left[\bS_{r,k}\right]_{ab}=\int \, w_{r,k}(\bk) e^{i\bk^T(\bx_a-\bx_b)}d\Omega_{\bk},
\end{equation}
where $\bx_a$ and $\bx_b$ are the location coordinates of the respective elements of the $k$-th RIS. The above integral is taken over all directions of $\bk/k_0$ on the unit sphere with differential solid angle $d\Omega_\bk$, and $w_{r,k}(\bk)$ is normalized so that, when $\bx_a=\bx_b$, the integral gives unity. The other correlation matrices in \eqref{eq:all_correlations} can be expressed in a similar way. Note that the generic weight function $w(\bk)$ can be characterized by the mean direction of arrival or departure $\bs_0$ (with $|\bs_0|=k_0$), and the Angle Spread (AS) $\sigma$ (in radians), so we can write:
\begin{equation}\label{eq:weight_fn_def}
    w(\bk)\propto e^{-\frac{|\bk-\bs_0|^2}{2\sigma^2k_0^2}}.
\end{equation}
It will, henceforth, be also convenient to express $w(\bk)$ in terms of $\bk_\parallel$, the component of $\bk$ parallel to the RIS and $k_z$, the component of $\bk$ perpendicular to the RIS (so that $\bk=(\bk_\parallel,k_\perp)$), as follows: $w(\bk_\parallel,k_z)$.  

%
%

\section{Capacity Analysis}
\label{sec:MI_Analysis}
In this paper, we are particularly interested in the limit where the RISs have large numbers of reflecting elements. However, in order to be able to obtain analytic expressions for the targeted MI performance metric, we will also consider the limit of large number of elements at the RISs, when the numbers of TX and RX antennas grow at the same rate with the reflecting elements.  Thus, in this section, we present novel asymptotic closed-form  expressions for the first two moments of the MI, which are nominally valid when the number of RIS elements, as well as the TX and RX antenna elements, become large at a fixed rate with each other. Nevertheless, as we shall see, our performance evaluation results are valid for realistic antenna and RIS sizes. We also discuss key properties 
of the correlation matrices involving the RISs, along with their physical interpretations and their implications on the optimization of the phase configurations.

\subsection{Mutual Information (MI) Statistics}
A key performance metric of communication links is the MI, which can be expressed for our considered system model in \eqref{eq:basic_channel_eq} as follows \cite{Foschini1998_BLAST1, Telatar1995_BLAST1}:
\begin{eqnarray}\label{eq:mut_info_def}
I \triangleq \log\det\left(\vec I_\nr + \rho\bG_{\rm tot} \bQ \bG_{\rm tot}^\dagger \right).
\end{eqnarray}
The above rate, expressed in nats per channel use, is achievable for Gaussian input signal vectors $\bx$ with covariance matrix $\bQ$, assuming that the RX knows the overall end-to-end channel matrix $\bG_{\rm tot}$ through pilot signaling \cite{hardware2020icassp,Swindlehurst_CE,HRIS_CE}. Since this channel matrix fluctuates due to fast-fading, the system's long-time performance is captured by the ergodic average of the MI, denoted by $\ex[I]$, where the expectation is taken over the fading distribution. It is evident in \eqref{eq:mut_info_def} that the value of the MI depends on all $\bPhi_k$'s. Therefore, in the next section, we will focus on optimizing those RIS phase configuration matrices, in order to maximize the ergodic MI for the considered RIS-empowered communication system, subject to statistical knowledge of the channel, which is more reliable and more conveniently acquired than the instantaneous one. When dealing with block-fading channels, the relevant communication performance metric is the outage MI, i.e., the mutual information achievable with a given outage probability \cite{Telatar1995_BLAST1}. In this case, we need additional information about the variability of the channel, which can be captured by the variance of the MI metric. In the following proposition, we summarize our asymptotic results regarding the ergodic MI and its variance.

\begin{Proposition}[Asymptotic Mean and Variance of MI]
\label{prop:ergMI} 
Let the channel matrix $\bG_{tot}$ be composed as in \eqref{eq:Gtot} with the matrices $\bG_d$, $\bG_{r,k}$, and $\bG_{t,k}$ for the direct channel and the outgoing and incoming channels from/to each $k$-th RIS be complex Gaussian random matrices with covariance given by \eqref{eq:Gd_cov}, \eqref{eq:Grk_cov}, and \eqref{eq:Gtk_cov}, respectively. In the limit $\nt, \nr, \ns\to\infty$ with fixed ratios $\beta_r\triangleq\nr/\nt$ and $\beta_s\triangleq\ns/\nt$, the ergodic MI per TX antenna element takes the following closed form:
\begin{align}\label{eq:S0}
 C \triangleq \frac{\ex[I]}{\nt}=& 
  \frac{1}{\nt}\sum_{k=1}^K \log\det \left(\bI_{\ns} +  t_{1k}r_{2k}\gamma_k \bSigma_k  \right)\nonumber
  \\ 
  &+\frac{1}{\nt}\log\det \left(\bI_\nr + \tilde{\bR}\right)\nonumber
  \\ 
  &+\frac{1}{\nt}\log\det \left(\bI_\nt + \rho\bQ\tilde{\bT}\right)
  \\ \nonumber
  &- r_d t_d-\sum_{k=1}^K\left(r_{1k}t_{1k}+r_{2k}t_{2k}\right),
\end{align}
while the MI's variance takes the limiting form:
\begin{align}\label{eq:Var(I)}
 {\rm Var}(I)\triangleq-\log\det({\bf \Lambda}),
\end{align}
where the matrices $\tilde{\bR}$, $\tilde{\bT}$, and $\bSigma_k$ are defined as: 
\begin{align}
    \tilde{\bR}&\triangleq r_d \bR_d + \sum_{k=1}^K r_{1k}\bR_k,\label{eq:R_tilde}\\
    \tilde{\bT}&\triangleq t_d\bT_d+\sum_{k=1}^K t_{2k}\bT_k,
    \label{eq:T_tilde}\\
    \bSigma_k &\triangleq \bS_{t,k}^{1/2}\bPhi_k^\dagger\bS_{r,k}\bPhi_k\bS_{t,k}^{1/2}
    \label{eq:Sigma_k_initial},
\end{align} 
the parameters $r_{1k}$, $t_{1k}$, $r_{2k}$, $t_{2k}$, $r_d$, and $t_d$ are the unique solutions of the following fixed-point equations:
\begin{equation}\label{eq:fp_eqs}
\begin{split}
    t_d &=\frac{1}{\nt}\Tr{(\bI_\nr+\tilde{\bR})^{-1}\bR_d},\\
    t_{1k} &=\frac{1}{\nt}\Tr{(\bI_\nr+\tilde{\bR})^{-1}\bR_k}, 
  \\ 
    r_d &= \frac{\rho}{\nt} \Tr{\bQ\bT_d\left(\bI_\nt + \rho\bQ\tilde{\bT}\right)^{-1}}, 
  \\ 
    r_{2k} &= \frac{\rho}{\nt} \Tr{\bQ\bT_{k}\left(\bI_\nt + \rho\bQ\tilde{\bT}\right)^{-1}}, 
  \\ 
   r_{1k}& = \frac{\gamma_kr_{2k}}{\nt} \Tr{\bSigma_k\left(\bI_{\ns} +
\gamma_kt_{1k}r_{2k}\bSigma_k \right)^{-1} }, 
  \\ 
   t_{2k} &= \frac{\gamma_kt_{1k}}{\nt} \Tr{\bSigma_k\left(\bI_{\ns} + \gamma_kt_{1k}r_{2k}\bSigma_k\right)^{-1} }.
\end{split}   
\end{equation}
and the $(4K+2)$-dimensional matrix ${\bf \Lambda}$ appears in \eqref{eq:Vmat_def}.
\end{Proposition}
\begin{proof}
The proof is delegated in the Appendix~\ref{app:proof_ergMI}.
\end{proof}
\begin{remark}[Decoupling of RISs]
A key simplification in the asymptotic limit is the decoupling of the phase configuration matrices $\bPhi_k$'s of the different RISs. As can be seen in \eqref{eq:S0}, these matrices appear in separate $\log\det(\cdot)$ terms.
\end{remark}
\begin{remark}[Central Limit Theorem for MI]
Following the discussion at the end of the Appendix~\ref{app:proof_ergMI}, it can be shown that the distribution of the MI performance converges weakly \cite{Hachem2006_GaussianCapacityKroneckerProduct} to a Gaussian, 
in the following sense:
\begin{equation}\label{eq:Gaussian_approximation}
    \lim_{\nt\to\infty}\frac{I-\nt C}{\sqrt{\log\det\left({\bf \Lambda}^{-1}\right)}}\sim {\cal N}(0,1),
\end{equation}
where the notation ${\cal N}(0,1)$ denotes the zero-mean and unit variance Gaussian distribution.
\end{remark}
\begin{remark}[Capacity Achieving $\bQ$]
Maximizing \eqref{eq:S0} over the input covariance matrix $\bQ$ with the constraint of a fixed trace, i.e., $\Tr{\bQ}=\nt$, will give the ergodic capacity per antenna of the system for fixed $\bPhi_k$'s.    
\end{remark}

\subsection{Asymptotic Properties of the Correlation Matrices}
\label{sed:Properties}

In this section, we will focus on the optimization of the ergodic MI with respect to all $\bPhi_k$'s. As it can be seen in \eqref{eq:S0}, this performance metric appears coupled to the correlation matrices $\bS_{r,k}$'s and $\bS_{t,k}$'s. It will be become apparent that, to be able to effectively optimize over each $\bPhi_k$, we need to take advantage of the structure of these correlation matrices. Specifically, we will use the fact that the RISs are positioned in (square) lattices. Thus, in this section, we will show that, in the limit of the  RISs' sizes increasing without bound, the eigenvectors of $\bS_{t,k}$'s and $\bS_{r,k}$'s essentially become Fourier modes, and their eigenvalues converge to the Fourier transforms of any line of the matrix. This was first suggested in \cite{Moustakas2000_BLAST1_new} and was later also discussed in the very recent papers \cite{Holographic_Pizzo,Holographic_WeiLi}. In the following proposition, we make this claim more concrete. 
We will make the technical assumption that the weight function $w(\bk_\parallel,k_z)/|k_z|$ is square-integrable. This constraint would only make a difference if a substantial portion of the relevant incoming/outgoing energy to/from the RISs originates from directions parallel to it. Furthermore, it should be noted that if we take into account the polarization properties of the metamaterial elements with polarization in the RIS plane, the latter assumption will be automatically satisfied.
For convenience, we henceforth drop the indices $t,r$, and $k$ from the correlation matrices $\bS_{t,k}$'s and $\bS_{r,k}$'s, and related quantities as $w_{t,k}$'s appearing in \eqref{eq:corr_mat_w(k)_def}, as well as their eigenvalues and eigenvectors.

\begin{figure}[!t]
	\centering
	\includegraphics[width=\columnwidth]{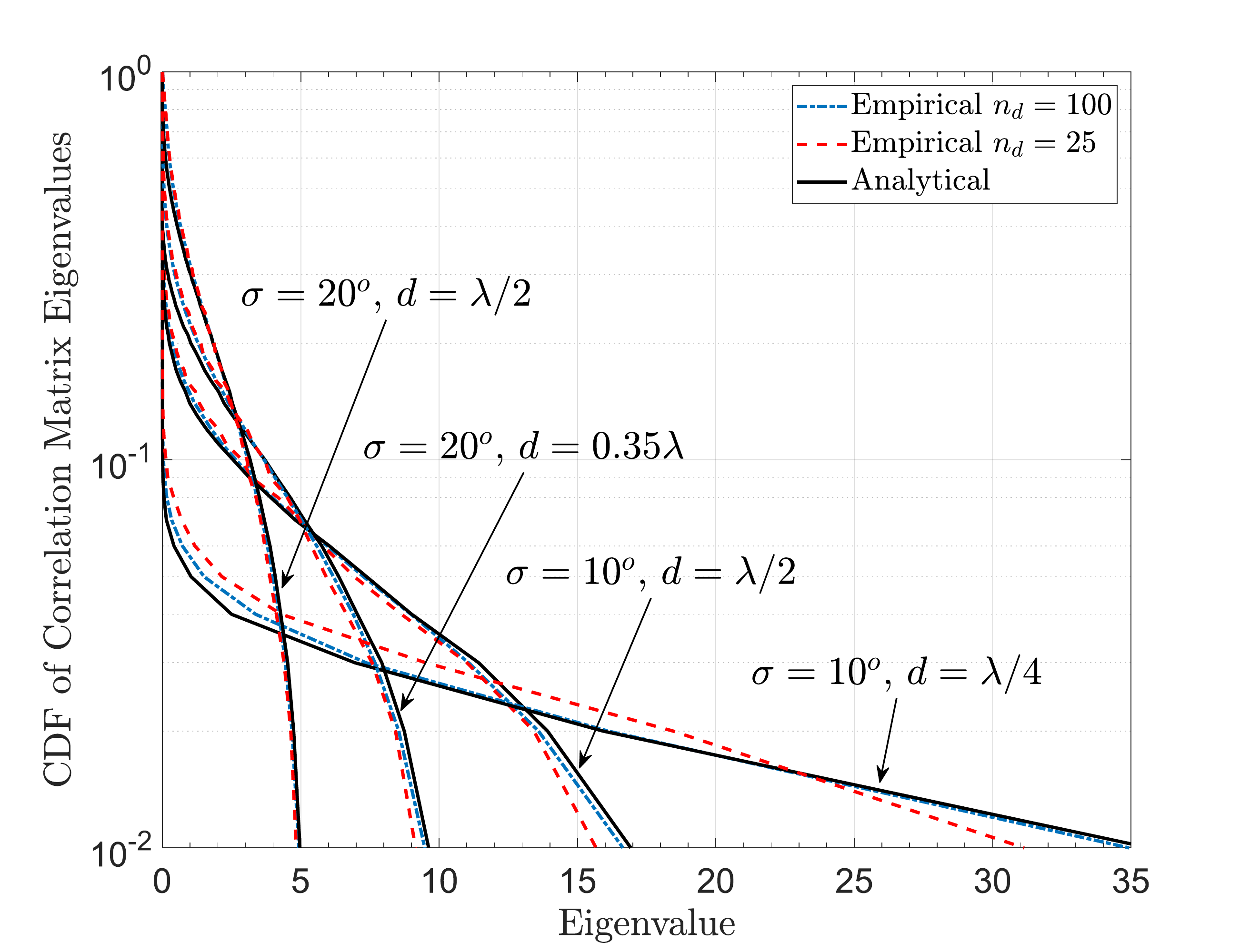}
	  \caption{The CDF of the eigenvalues of the correlation matrices $\bS_{t,k}$ and $\bS_{r,k}$. The theoretical curves are based on Proposition~\ref{prop:eigenvalues} for the Gaussian weight function $w(\bk)$ in \eqref{eq:weight_fn_def} with $\mathbf{s}_0=k_0\hat{\be}_\perp$, where $\hat{\be}_\perp$ is the incoming EM wave direction vertical to the $k$-th RIS. The maximum eigenvalue for each case increases when the inter-element distance $a$ (as can be directly seen in \eqref{eq:prop:eigenvalue}) and/or the AS $\sigma$ decrease. The agreement between the analytical results and the corresponding Monte Carlo simulations is shown to be very good. 
	  }
		\label{fig:cdf_corr_mat_eigs}
\end{figure}

\begin{Proposition}[Limiting Behavior of Correlation Matrices]
\label{prop:eigenvalues}
Let the reconfigurable reflecting elements of an RIS form a square grid located on the lattice $\bx=\bn a$, where $a$
denotes the RIS inter-element spacing and $\bn\triangleq(n_1,n_2)$ is a two-dimensional integer vector with $n_1,n_2=1,2,\ldots,\nd\triangleq \sqrt{\ns}$, so that any $(i,j)$-element of the correlation matrix $\bS$ can be obtained in terms of a single function $S(\bn)$, with $\bn\in\ZZ^2$, as follows:
\begin{equation}\label{eq:S(n)_def}
    \left[\bS\right]_{i,j}=S(\bn_i-\bn_j)\triangleq\int w(\bk_\parallel,k_z)\,\, e^{i\bk^T(\bn_i-\bn_j)a}d\Omega_{\bk},
\end{equation}
where $\bn_i a$ and $\bn_j a$ are the location coordinates of the respective elements of the RIS. We further assume  that $w(\bq,\pm k_z(\bq))/k_z(\bq)$ is square integrable in $\bq\in \left[-\frac{\pi}{a},\frac{\pi}{a}\right]^2$, where $k_z(\bq)\triangleq\sqrt{k_0^2-|\bq|^2}$ for $k_0=2\pi/\lambda\leq \pi/a$. Then, the Fourier transform of $S(\bn)$ is given by the expression:
\begin{equation}\label{eq:prop:eigenvalue}
\eta(\bq)\triangleq \left(\frac{\lambda}{a}\right)^2\frac{w(\bq,k_z(\bq))+w(\bq,-k_z(\bq))}{\sqrt{1-\frac{|\bq|^2}{k_0^2}}} \Theta(k_0-|\bq|)
\end{equation}
for $\bq\in\left[-\frac{\pi}{a},\frac{\pi}{a}\right]^2$, where $\Theta(x)$ is the step function with $\Theta(x)=1$ if $x>0$ and zero if $x<0$. Then, in the limit $\ns\to\infty$, the eigenvalues of $\bS$ are distributed as $\eta(\bq)$ \cite{Tyrtyshnikov1996_UnifyingApproachToeplitz}.

We next define the block-circulant matrix $\bC$ with elements $[\bC]_{i,j}\triangleq S((\bn_i-\bn_j)({\rm mod}\text{ }n_d))$, where 
$\bn({\rm mod}\text{ }\nd)\triangleq (n_1({\rm mod}\text{ }n_d),n_2({\rm mod}\text{ }\nd))$. This matrix has the Fourier modes $\bu(\bq_\bm)$ as eigenvectors, where $\bq_\bm$ are a discretized and equidistant set of wave-vectors defined as $\bq_\bm\triangleq2\pi \bm/(\nd a)$ with $\bm\triangleq(m_1,m_2)$ for $m_1,m_2=1,2,\ldots,\nd$. The elements of these Fourier modes are given $\forall$$j=1,2,\ldots,\nd$ as follows: 
\begin{equation}\label{eq:Fmode}
    \left[\bu(\bq_\bm)\right]_j = \frac{e^{i\bq_\bm^T\bn_j a}}{\sqrt{\ns}}.
\end{equation}
Then $\bS$ and $\bC$ are asymptotically equivalent in the sense that:
\begin{equation}\label{eq:|S-C|^2}
\lim_{\ns\to \infty} \frac{1}{\ns}\left\Vert \bS-\bC\right\Vert^2_F = 0.
\end{equation}
\end{Proposition}
\begin{proof}
To prove \eqref{eq:prop:eigenvalue} we start from \eqref{eq:S(n)_def},  multiply with $e^{-i\bq^T\bm a}$, and sum over $\bm\in \ZZ^2$, yielding:
\begin{align}\label{eq:prop:eigenvalue1}
\eta(\bq)&=\sum_{\bm\in\ZZ^2}
e^{-i\bq^T\bm a} \int w(\bk_\parallel,k_z)\,\, e^{i\bk^T \bm a} d\Omega_{\bk}\nonumber
\\ \nonumber
&=\int \frac{\delta(k-k_0)}{4\pi k_0} w(\bk_\parallel,k_z) \sum_{\bm\in \ZZ^2}
e^{-i(\bk-\bq)^T\bm a}d\bk
\\ 
&=\int \frac{\delta(k-k_0)}{4\pi k_0} w(\bk_\parallel,k_z)
\\ \nonumber
&\hspace{0.8cm}\times\left(\frac{2\pi}{a}\right)^2\sum_{\bp\in\ZZ^2}
\delta(\bk_\parallel-\bq-\frac{2\pi\bp}{a}) d\bk
\\ \nonumber
&=
\left(\frac{\lambda}{a}\right)^2\frac{w(\bq,|k_z(\bq)|)+w(\bq,-|k_z(\bq)|)}{\frac{|k_z(\bq)|}{k_0}} \Theta(k_0-|\bq|).
\end{align}
The second equality in this expression follows from the introduction of a Dirac $\delta$-function over the norm $|\bk|=k$, together with the corresponding integral over $k$. The third equality follows from the Poisson summation formula, while the last equality follows from the change of variable inside the Dirac $\delta$-function, which can be summarized with the following identity: 
\begin{equation}
    \delta(|{\bf k}|-k_0)\!=\!\frac{2k_0\left(\delta(k_z\!-\!\sqrt{k_0^2\!-\!|\bk_\parallel|^2})\!+\!\delta(k_z+\sqrt{k_0^2\!-\!|\bk_\parallel|^2})\right)}{\sqrt{k_0^2\!-\!|\bk_\parallel|^2}}.
\end{equation}

Since, by assumption, $w(\bq,\pm k_z(\bq))/k_z(\bq)$ is  square integrable, Theorem 8.2 in \cite{Tyrtyshnikov1996_UnifyingApproachToeplitz} can be applied to show that the eigenvalues of $\bS$ are asymptotically distributed as $\eta(\bq)$. Furthermore, under the same assumptions, Theorem 7.1 in \cite{Tyrtyshnikov1996_UnifyingApproachToeplitz} can be applied to prove \eqref{eq:|S-C|^2}.
\end{proof}
There are a number of conclusions we can obtain from the latter proposition; these are summarized in the sequel.
\begin{remark}[Eigenvalues of Correlation Matrices]
When the correlation matrices $\bS_{t,k}$'s and $\bS_{r,k}$'s become large in terms of their element numbers, their eigenvalues asymptotically behave as $\eta_{t,k}(\bq)$'s and $\eta_{r,k}(\bq)$'s, respectively, and are parametrized by the wavevector $\bq\in\left[-\frac{\pi}{a},\frac{\pi}{a}\right]^2$, which is in the plane of each corresponding $k$-th RIS. In addition, we see that eigenvalues with wave-vectors $|\bq|>k_0$ are asymptotically zero. Hence, comparing the area of the non-zero eigenvalues to the total support of eigenvalues, the proportion of the non-zero eigenvalues is roughly equal to $\pi k_0^2/(2\pi/a)^2$. In contrast, due to the dependence of the eigenvalues on the weight function, when the AS $\sigma$ is relatively small, the weight function $w_{t,k}(\bk)$ defined in \eqref{eq:weight_fn_def} (and similarly $w_{r,k}(\bk)$), and therefore the corresponding eigenvalue $\eta_{t,k}$ (and similarly $\eta_{r,k}$), is negligible outside the region $|\bk-\bs_0|\leq \frac{k_0\sigma}{2\pi}$. Thus, we may estimate the number of non-negligible eigenvalues to be roughly $\frac{\sigma^2 a^2 \ns}{(2\pi)^2\lambda^2}$. In Fig.~\ref{fig:cdf_corr_mat_eigs}, we plot the Cumulative Distribution Function (CDF) of the eigenvalues for various AS values $\sigma$ and RIS inter-element spacings $a$. We can observe very good agreement of the theoretical formula in Proposition~\ref{prop:eigenvalues} for the eigenvalues with the numerical diagonalization of the correlation matrices. 
\end{remark}
\begin{remark}[Behavior of Eigenvectors of Correlation Matrices]
As a result of the closeness of the matrices $\bS_{t,k}$'s (and $\bS_{r,k}$'s) to their corresponding block-circulant matrices, as shown in \eqref{eq:|S-C|^2}, the eigenvector with eigenvalue parametrized with the wavevector $\bq$ will be approximately equal to  the Fourier mode $\bu(\bq)$ defined in \eqref{eq:Fmode}. Combining this observation with the fact that there are no non-zero eigenvalues with $|\bq|>k_0$, we see that each incoming wavevector $\bk$, with $|\bk|=k_0$, corresponds to the eigenvector which has the same projection $\bq=\bk_\parallel$ on the plane of the RIS and a vertical component given by $\pm k_z(\bq)$. This observation showcases the relationship of this result with the so-called ``holographic'' concept introduced recently in the wireless literature \cite{huang2019holographic,Holographic_Pizzo, Holographic_WeiLi}, according to which the EM modes of a two-dimensional surface are related with the incoming (or outgoing) plane waves onto the surface.
\end{remark}

\section{Capacity Optimization}\label{sec:MI_Optimization}
The aim of this section is the optimization of the ergodic MI  with respect to the phase configurations and the placements of the RISs, i.e., the elements of $\bPhi_k$'s and the distance parameters $\{d_{k,1},d_{k,2}\}_{k=1}^K$, respectively. Following the system model in \eqref{eq:Gtot} and the geometry appearing in Fig. \ref{fig:system_model} with the distance-dependent pathloss given by \eqref{eq:gamma}, and using the closed-form asymptotic MI expression in \eqref{eq:S0}, our Optimization Problem (OP) formulation is mathematically expressed as follows\footnote{Note that, in practice, and in order to reduce the search space, there should be constraints on the placements of the RISs, and consequently on the optimization variables $d_{k,1}$'s and $d_{k,2}$'s, due to deployment restrictions and the double-pathloss effect \cite{bjornson_intelligent_2020}. }:
\begin{align*} 
\begin{split}
    \mathcal{OP}_1: \,\, \max_{\substack{\{\bPhi_k\}_{k=1}^K,\\\{d_{k,1},d_{k,2}\}_{k=1}^K}} \,\,&C\left(\{\bPhi_k\}_{k=1}^K,\{d_{k,1},d_{k,2}\}_{k=1}^K\right) 
    \\    \hspace{0.4cm}\text{s.t.} \quad &\lvert [\bPhi_k]_{n,n} \rvert = 1  \, \, \forall k,n.
\end{split}
\end{align*}
To perform the optimization, we will start by keeping $d_{k,1}$'s and $d_{k,2}$'s fixed to first find the $\bPhi_k$'s that maximize the ergodic MI, and subsequently we will solve $\mathcal{OP}_1$ with respect to $d_{k,1}$'s and $d_{k,2}$'s using the latter optimized RIS phase configurations.  

Before moving on, it should be pointed out that, in the large system limit, the above OP can be applied to both the ergodic and the outage capacity optimizations. This happens because, in this limit, as can be seen in Proposition \ref{prop:ergMI}, the variance of the MI performance per transmit antenna is ${\rm Var}(I)/\nt^2$, and therefore, it vanishes.

\subsection{Optimization w.r.t. $\bPhi_k$'s}
As mentioned above, we start by fixing the positions of the RISs and maximize the ergodic MI with respect to the phase-matrix $\bPhi_k$ of each $k$-th RIS. However, we observe that the asymptotic expression of the ergodic MI performance appearing in \eqref{eq:S0}, \eqref{eq:Sigma_k_initial}, and \eqref{eq:fp_eqs} has a key simplification compared to \eqref{eq:mut_info_def}; namely, as shown in \eqref{eq:S0}, the dependence on the reflection matrices $\bPhi_k$'s decouples into a sum of logarithms. Although the decoupling is only partial, since there is some dependence through the fixed-point equation parameters in \eqref{eq:fp_eqs}, this implies that our intended joint optimization of $\bPhi_k$'s can be done separately\footnote{It should be stressed, however, that this is a consequence of the statistical independence of the channel matrices for each RIS. If those channels were correlated, this decoupling would no longer be possible.}. Based on this observation, we may proceed by adopting an iterative approach based on alternating optimization \cite{J:alternating_minimization}. 

At each algorithmic iteration, we  
first consider the variables $r_{1k}$, $t_{1k}$, $r_{2k}$, $t_{2k}$, $r_d$, and $t_d$ as fixed and optimize the ergodic MI with respect to the matrices $\bPhi_k$'s. This optimization procedure may still be broken into independent optimizations over each $\bPhi_k$ separately, due to the fact that these matrices appear in independent logarithms in \eqref{eq:S0}. Hence, the multi-RIS phase configuration design problem can be simplified as follows:  
\begin{align*} 
\begin{split}
    \mathcal{OP}_2: \,\,\sum_{k=1}^K \max_{\{\bPhi_k\}_{k=1}^K} &\log\det \left(\bI_{\ns} +  
t_{1k}r_{2k}\gamma_k\bPhi_k^\dagger\bS_{r,k}\bPhi_k\bS_{t,k}  \right)
\\    \hspace{0.4cm}
\text{s.t.} \quad &\lvert [\bPhi_k]_{n,n} \rvert = 1  \, \, \forall k,n.
\end{split}
\end{align*}
Afterwards, we substitute the $\mathcal{OP}_2$ solution into \eqref{eq:fp_eqs} to calculate $r_{1k}$, $t_{1k}$, $r_{2k}$, $t_{2k}$, $r_d$, and $t_d$. The latter two steps are repeated at each iteration until convergence, or when reaching a threshold value for the ergodic MI objective. 

We will next consider two independent ways to obtain the optimum $\bPhi_k$'s: a numerical optimization approach based on conventional techniques, and a novel analytic optimization based on the findings in the previous Section~\ref{sec:MI_Analysis}.

\subsubsection{Numerical Solution}\label{sec:Numerical_Solution}
In the first numerical approach, the $\bPhi_k$'s solving $\mathcal{OP}_2$ can be computed via the algorithm of either \cite{Zhang_Capacity} (algorithmic iterations per each RIS element) or \cite{PLS_Kostas} (iterations per each RIS phase configuration matrix). We have adopted the former approach in our numerical evaluation, however, the detailed algorithm is omitted here due to space limitations.

\subsubsection{Analytic Solution}
\label{sec:Analytical_Solution}  
Given the complexity of the above numerical algorithm, it is advisable to look for a simpler solution to the problem of optimizing the $\bPhi_k$'s, at least in limiting cases. We start by noting that each $\bPhi_k$ enters the ergodic MI calculation only through the matrix $\bSigma_k$, which itself depends on the correlation matrices $\bS_{t,k}$ and $\bS_{r,k}$. Therefore, capitalizing on Proposition \ref{prop:eigenvalues}, we will use the properties of the limiting structure of the correlation matrices $\bS_{r,k}$'s and $\bS_{t,k}$'s to propose a matrix $\bPhi_k$, which we can show to be optimal for small values of the ASs. 

To this end, we first express each $\bSigma_k$ in terms of the eigenvalues and eigenvectors of $\bS_{t,k}$ (denoted as $\{\eta_{tk,n}\}_{n=1}^\ns$ and $\{\bu_{tk,n}\}_{n=1}^\ns$) and $\bS_{r,k}$ (denoted as $\{\eta_{rk,n}\}_{n=1}^\ns$ and $\{\bu_{rk,n}\}_{n=1}^\ns$) as follows:
\begin{align}
\label{eq:Sigma_k}
\bSigma_k&=\sum_{\ell,\ell'=1}^\ns \bu_{rk,\ell}\bu^\dagger_{rk,\ell'} \sqrt{\eta_{rk,\ell}\eta_{rk,\ell'}} \sum_{m=1}^\ns\eta_{tk,m} \kappa_{k,m\ell} \kappa_{k,m\ell'}^*,
\end{align}
where we have used the notation:
\begin{align}
\label{eq:alpha_k}
\kappa_{k,m\ell}&\triangleq\bu_{tk,m}^\dagger\bPhi_k\bu_{rk,\ell}.
\end{align}
The above expression highlights that the effect of each $\bPhi_k$ on the MI is filtered through the eigenvectors of $\bS_{t,k}$'s and $\bS_{r,k}$'s in the form expressed by $\kappa_{k,m\ell}$'s. 

To obtain some intuition on the above expressions, we analyze two limiting cases for the ASs of the correlation matrices. In the first case, the AS at each $k$-th RIS is assumed very high, making $\bS_{t,k}$ and $\bS_{r,k}$ essentially proportional to the unit matrix. In this case, the optimization over $\bPhi_k$'s, as expressed in $\mathcal{OP}_2$, is immaterial, since $\bSigma_k\approx\bPhi_k^\dagger\bPhi_k=\bI_\ns$. The opposite case is more instructive, namely when the AS is very small so that $\bS_{t,k}$ and $\bS_{r,k}$ are unit-rank matrices, corresponding effectively to two Line-Of-Sight (LOS) channels. In this case, it holds $\forall a,b$ that $[\bS_{r,k}]_{a,b}\approx\ns[\bu_{rk,1}\bu_{rk,1}^\dagger]_{a,b}=e^{i\bq_{rk,1}^T(\bx_a-\bx_b)}$ and $[\bS_{t,k}]_{a,b}=\ns [\bu_{tk,1}\bu_{tk,1}^\dagger]_{a,b}=e^{i\bq_{tk,1}^T(\bx_a-\bx_b)}$, where the wavevectors $\bq_{tk,1}$ and $\bq_{rk,1}$ correspond to the projection on the surface of each $k$-th RIS of the mean direction of arrival and departure of the EM wave's energy. This yields the following simplified expression for   $\bSigma_k$:  
\begin{align}\label{eq:unit_rank_Sigma}
    \bSigma_k= \ns |\kappa_{k,11}|^2 \bu_{rk,1}\bu^\dagger_{rk,1},
\end{align}
where $\kappa_{k,11}$ is given using \eqref{eq:alpha_k} by:
\begin{align}
\label{eq:alpha_k_new}
\kappa_{k,11}&=\bu_{rk,1}^\dagger\bPhi_k\bu_{tk,1}=\frac{1}{\ns}\sum_{n=1}^{\ns} e^{i\phi_{k,n}}e^{-i(\bq_{tk,1}-\bq_{rk,1})^T\bx_n}.
\end{align}
The above expression implies that the $\bPhi_k$'s maximizing the ergodic MI are such that $\phi_{k,n}=(\bq_{tk,1}-\bq_{rk,1})^T\bx_n$ $\forall$$k,n$. We can clearly see that, in the case of geometrical optics, for which the components of the incoming and outgoing direction vectors parallel to each $k$-th RIS are equal (i.e., $\bq_{tk,1}=\bq_{rk,1}$), no phase optimization is necessary. However, when this is not the case, the phase configuration optimization of the multiple RISs will produce significant gains. Interestingly, the Fourier form of the eigenvectors will allow us to gain insight on the optimization process and obtain the optimal $\bPhi_k$'s in $\mathcal{OP}_2$ in closed form. Indeed, when the eigenvalue distributions of $\bS_{r,k}$'s and $\bS_{t,k}$'s are a displacement of one another in the domain space of 
$\bq$, as seen in the insert figure of Fig.~\ref{fig:MI_AS_124RIS}, the differences of the $\bq$-vectors of the corresponding eigenvalues in each distribution are constant and equal to the difference between the $\bq$-vectors of the maximum eigenvalues of the matrices, i.e., $\bq_{tk,\ell}-\bq_{rk,\ell}=\bq_{tk,1}-\bq_{rk,1}$ $\forall$$\ell\neq1$. Thus, setting the phases of all $\bPhi_k$'s for the $K$ RISs as follows:
\begin{equation}\label{eq:Phi_opt}
e^{i\phi_{k,n}}=e^{i(\bq_{1t}-\bq_{1r})^T\bx_n}\,\,\forall k,n,   
\end{equation}
resulting using \eqref{eq:Sigma_k} in $\kappa_{m\ell,k}=\delta_{m,\ell}$ ($\delta_{m,\ell}$ is the Kronecker delta function) $\forall$$m,\ell=1,2,\ldots,N_s$,
is optimal. Clearly, if the domain of non-negligible eigenvalues of each $\bS_{t,k}$ and $\bS_{r,k}$ are not related by a shift in $\bq$, then the above conjecture for the phase configuration of the RIS elements may not be optimal, but will be close, and thus, can serve as an initial condition for further optimization. 

The analysis above clarifies the relationship between the ``holographic'' EM modes of the RIS and the tunable phases of its elements \cite{Holographic_Pizzo, Holographic_WeiLi}. Specifically, these phases need to be related with phase differences (or correspondingly wave-modes) between the incoming and outgoing EM waves. 

\subsection{Optimization w.r.t. $d_{k,1}$'s and $d_{k,2}$'s}
We will now discuss the placement optimization of the RISs in space, thereby optimizing the distance of each $k$-th RIS from the TX, namely the parameter $d_{k,1}$, and from the RX, i.e., $d_{k,2}$. To keep the optimization space tractable, we will keep the location of each RIS on a straight line, parallel to the line connecting the TX-RX arrays, following the example illustrated in Fig$.$~\ref{fig:system_model}, leaving more general setups for future investigation. 

The optimum placement of the $1$-st RIS for a given TX-RX distance $d$ can be obtained from the following proposition; it similarly holds for the placement of each $\ell$-th RIS with $\ell=2,3,\ldots,K$.

\begin{Proposition}[Pathloss Behavior of RISs]
\label{prop:OptSNR}
The SNR of the signal due to the $1$-st RIS, $\gamma_1$, appearing in expression \eqref{eq:gamma} for the example communication system in Fig$.$~\ref{fig:system_model}, and its corresponding 2D Cartesian coordinate system, is maximized when the RIS is placed in the midpoint of the line segment ${\rm T}_1{\rm R}_1$ (i.e., at the $(0,h)$ point with $|h|$ being the length of the line segment ${\rm T}_1{\rm T}$) when $h\geq\frac{d}{2}$, or at the points $(\pm\sqrt{\frac{d^2}{4}-h^2},h)$ when $h<\frac{d}{2}$. 
\end{Proposition}
\begin{proof}
From the Pythagorean theorem in Fig. \ref{fig:system_model} we have that $d_{1,1}^2=h^2+\left(\frac{d}{2}-x\right)^2$ and $d_{1,2}^2=h^2+\left(\frac{d}{2}+x\right)^2$, yielding:
\begin{equation}\label{eq:product}
d_{1,1}^2d_{1,2}^2=\left(h^2+\frac{d^2}{4}+x^2\right)^2-d^2x^2.
\end{equation} 
Clearly, the value of $x$ that minimizes \eqref{eq:product}, let it be $x^*$, results in the maximization of $\gamma_1$ in \eqref{eq:gamma}. The left-hand side of the above expression is a quadratic function of $x^2\geq 0$, and exhibits its minimum value when $(x^{*})^2=\max(\frac{d^2}{4}-h^2,0)$. Hence, when $h\leq\pm\frac{d}{2}$, there are two equivalent minima at $x^*=\pm \sqrt{\frac{d^2}{4}-h^2}$, while $x^*=0$ otherwise. The latter case implies that the RIS should be placed in the midpoint of the line segment ${\rm T}_1{\rm R}_1$.   
\end{proof}

The above analysis complements the relevant literature (e.g., \cite{bjornson_intelligent_2020,Optimal_Placement}), identifying different regimes that depend on the distance of each RIS from the TX-RX line segment. Indeed, the above proposition shows that when the vertical distance $h$ is larger than $d/2$, the optimal location for each RIS is the midpoint between TX and RX. In contrast, when this vertical distance is reduced from the above value, the optimal location splits into two equivalent symmetrical locations that move continuously towards the TX and RX locations, respectively. In the limit that $h$ vanishes (i.e., $h\rightarrow0$), the SNR becomes infinite, which is, of course, physically impossible, since the considered far-field approximation is no longer valid. Nevertheless, the analysis indicates that, whenever possible, each RIS should be placed close to the TX or RX antenna arrays, playing the role of a reflector. In the limiting case, either the TX or RX can be equipped with an RIS-based front-end (a.k.a. the holographic MIMO surface paradigm \cite{DMA_2020,Holographic_Pizzo, Holographic_WeiLi, Beyond_Massive_MIMO}), or the RIS can be deployed at either of the them as lens \cite{RIS_lens}, operating in the near-field of a conventional TX or RX array. Of course, a different analysis is necessary for such a geometry.

\section{Numerical Results and Discussion}\label{sec:Numerical_Results}
\begin{figure}[!t]
	\centering
	\includegraphics[width=\columnwidth]{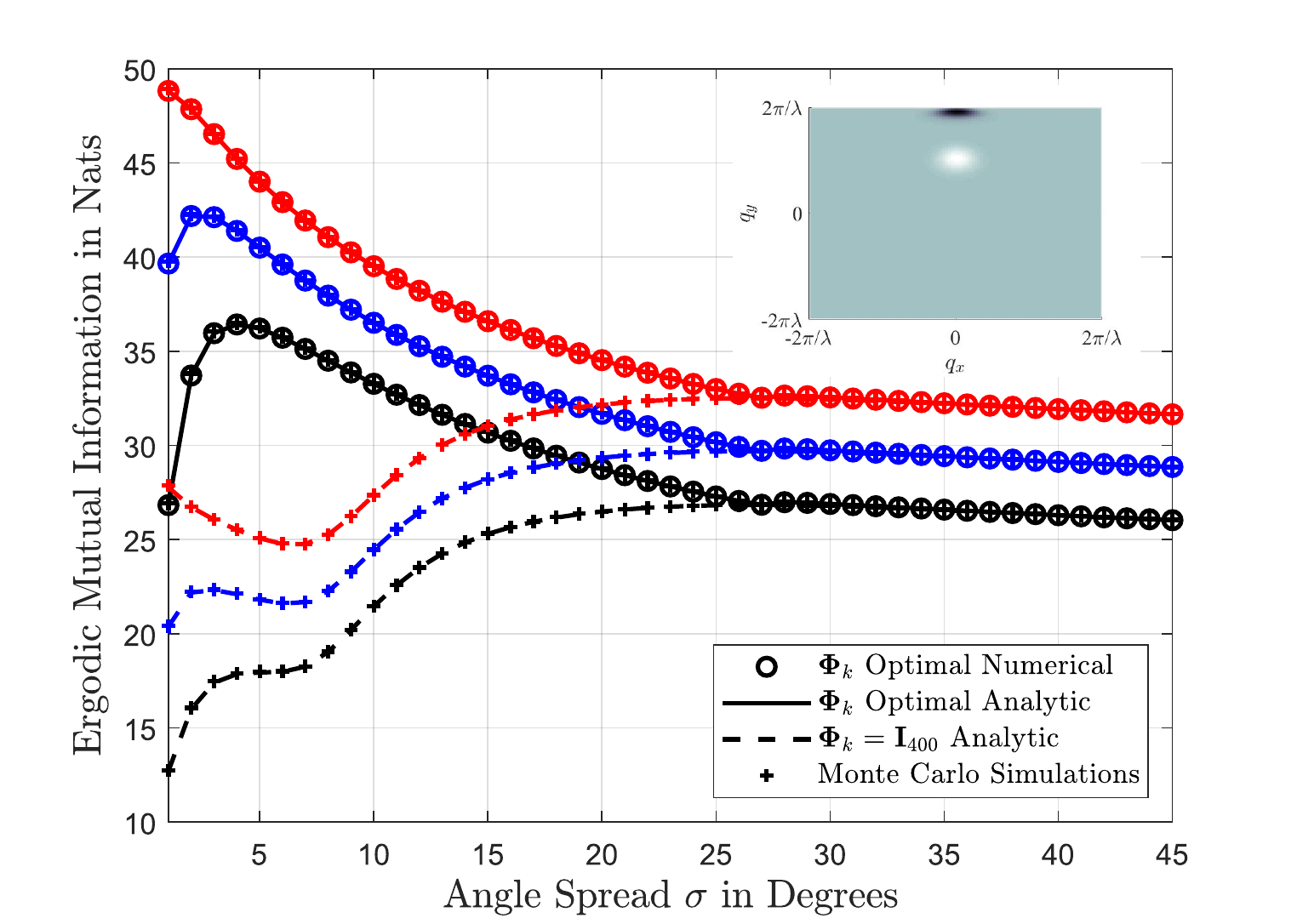}
	  \caption{The ergodic MI performance in nats per channel use for $\nt=8$, $\nr=4$, and $\rho=10$ dB versus the AS $\sigma$ in degrees, for a system operating at $2.5$ GHz with $K=1$ (black), $K=2$ (blue), and $K=4$ (red) RISs present. Each RIS is a square lattice of $20\times 20$ reflecting elements of inter-element distance $\lambda/2=6$ cm, and the case where there is no direct link between TX and RX has been considered. As depicted, RIS optimization plays a prominent role in low AS values for the simulated large $\ns=400$ value (i.e., correlated channels), while for large ASs, the RIS optimization becomes unnecessary. The inset figure depicts the distribution of the eigenvalues of the incoming (white ellipsis) and outgoing (black ellipsis) EM waves with incoming angle $\theta_1=30^o$ and outgoing angle $\theta_2=70^o$, respectively, as well as AS $\sigma=5^o$.}
		\label{fig:MI_AS_124RIS}
\end{figure}
\begin{figure}[!t]
\centering
\includegraphics[width=1.02\columnwidth]{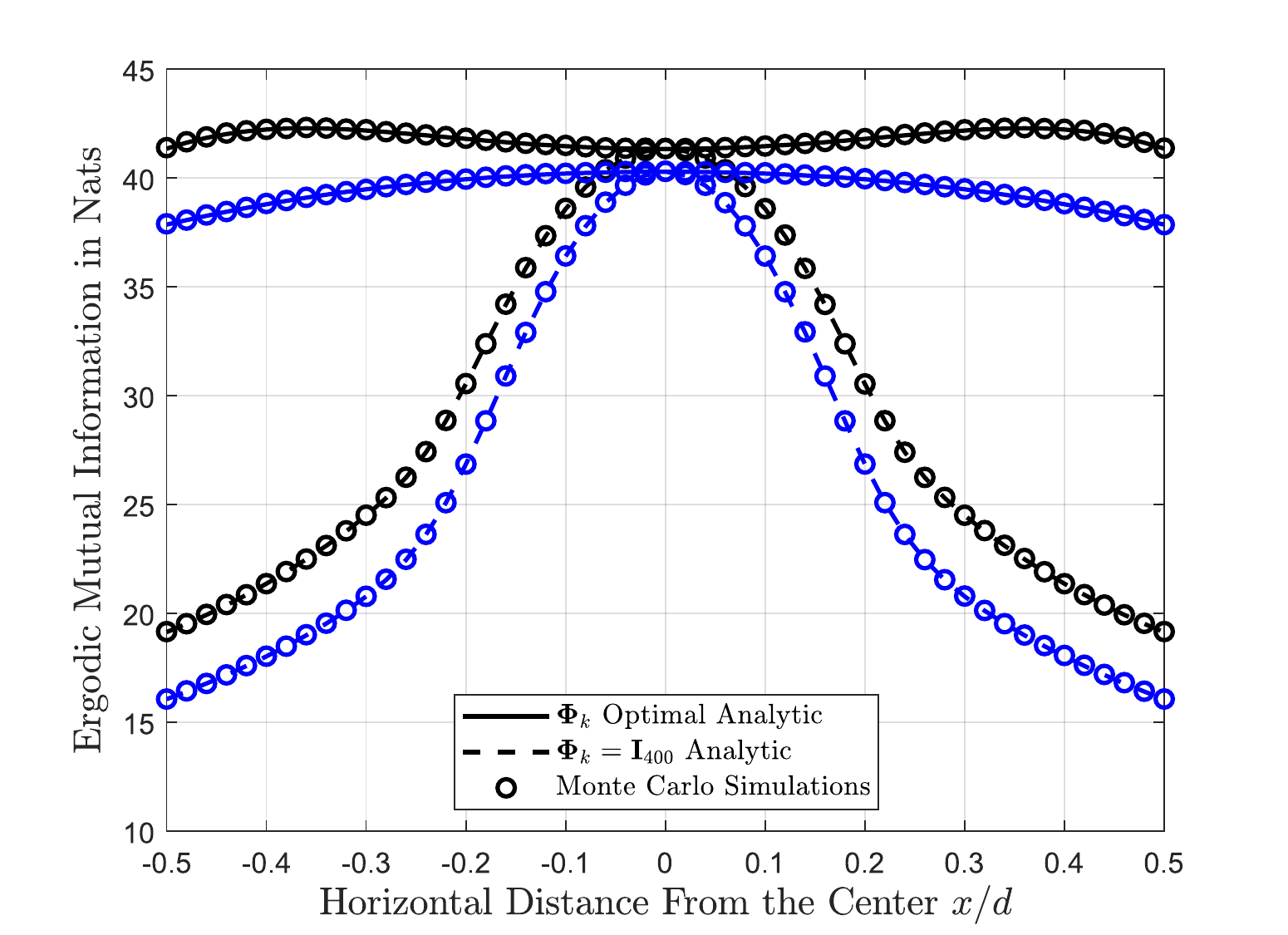}
\caption{The ergodic MI performance in nats per channel use, when $K=2$ RISs are present, as a function of the normalized horizontal distance $x/d$ (unitless) between the TX and RX antenna arrays measured from the origin in Fig$.$~\ref{fig:system_model}. The SNR between the latter end nodes varies with the horizontal distance $x$ as shown in \eqref{eq:gamma} for $h=0.7d$ (blue curves) and with $h=0.3d$ (black curves). The actual variation of the SNR with $x$ for these two cases is shown in the following Fig.~\ref{fig:SNR}. The dashed curves represent the case where $\bPhi_k=\bI_{400}$ for $k=1$ and $2$, while the solid curves correspond to the case where $\bPhi_k$'s are optimized using the analytical expression \eqref{eq:Phi_opt}, for the corresponding mean angles of arrival and departure that are mapped to the specific location of the RISs (see Fig. \ref{fig:system_model}). In the case of $h=0.7d$ (blue), for which the SNR is maximum at the center, the MI gets clearly its maximum value at this point $x=0$. In contrast, in the case of $h=0.3d$, for which the SNR becomes maximum away from the center (see the inset figure and Proposition~\ref{prop:OptSNR}), the maximum MI appears at the point where the SNR is maximized. Nevertheless, it is evident that the optimization of the RIS makes the MI remain nearly constant as a function of distance, despite the variations of both the SNR and the angles of arrival and departure of the incoming and outgoing EM waves, respectively, at the RISs.}
\label{fig:MI_x_2RIS}
\end{figure}

In this section, we present numerically evaluated results on the statistics of the MI performance for the considered multi-RIS-empowered wireless communications system between a multi-antenna TX-RX pair, using both the analytical (``Optimal Analytic'') and numerical (``Optimal Numerical'' using \cite{Zhang_Capacity}) optimization approaches presented in the previous sections, as well as Monte Carlo simulations. For simplicity, in all performance evaluation figures that follow, we have not included the direct path $\bG_d$ (considering it either blocked or highly attenuated) and assumed that both the TX and RX antenna arrays exhibit uncorrelated fading. We also showcase the accuracy of our analytic results for the mean and variance of the MI by comparing them with Monte-Carlo generated channel matrix instantiations. In all but Fig. \ref{fig:MI_x_2RIS}, we have assumed that the RISs are located at the midpoint between TX-RX, thus, we have set $\gamma_k=1$ $\forall$$k$.
\begin{figure}[!t]
	\centering
	\includegraphics[width=\columnwidth]{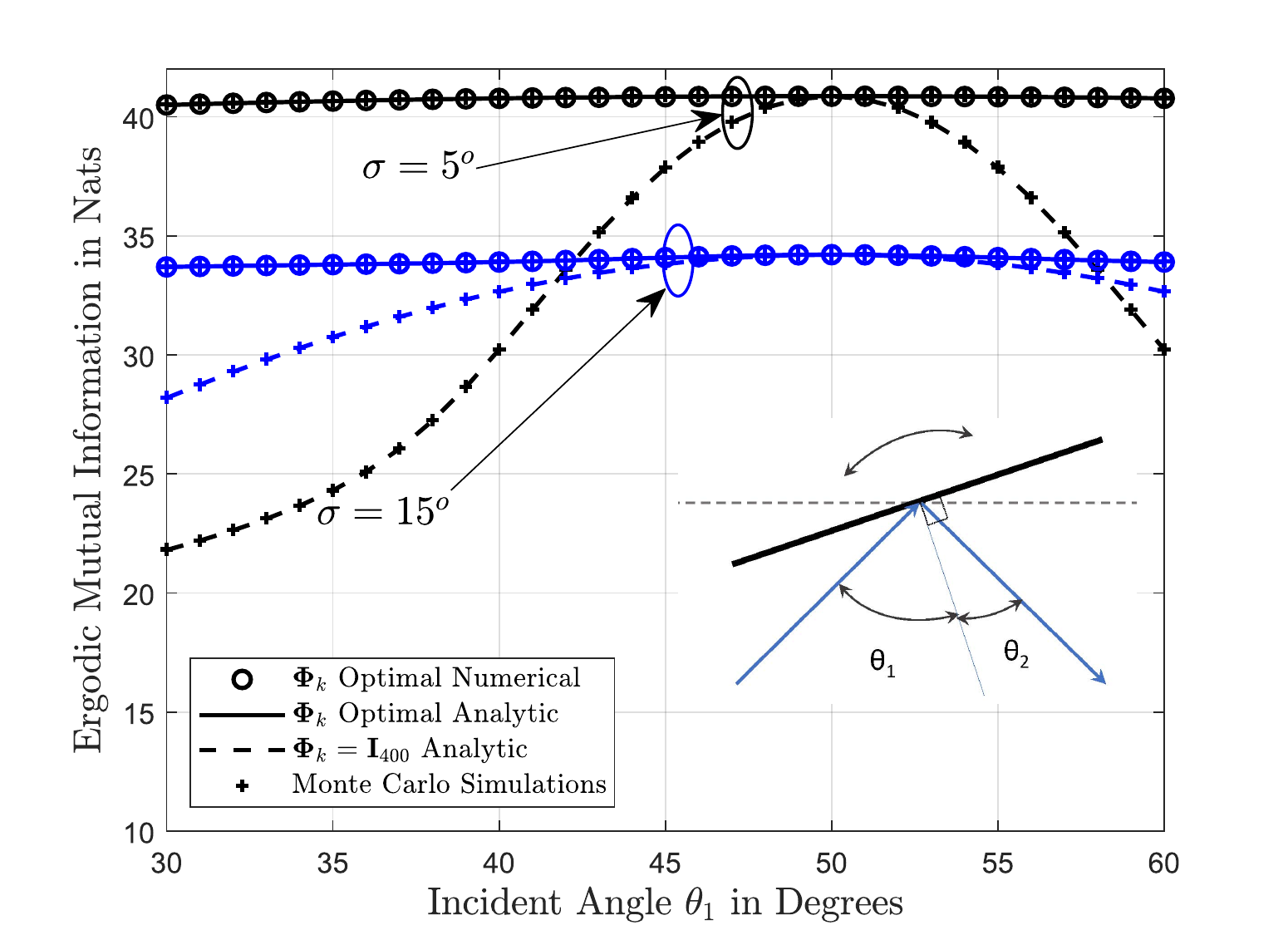}
\caption{The ergodic MI performance in nats per channel use in the presence of $K=2$ RISs as a function of the mean incoming angle $\theta_1$ in degrees (essentially as a function of the degree of tilting the RISs), considering the two different values $\sigma=5^o$ and $15^o$ of the incoming and outgoing EM waves' ASs. The RISs are equidistant from the TX and RX antennas, with the total mean incident and outgoing angles satisfying the relationship $\theta_{\rm tot}=\theta_1+\theta_2=100^o$, and are tilted with respect to the horizontal axis, as shown in the inset figure. The rest of the system parameters are the same with Fig$.$~\ref{fig:MI_AS_124RIS}. The figure showcases that, in the low AS case for fixed phase configurations $\bPhi_k=\bI_{400}$ for $k=1$ and $2$, the maximum MI is achieved when $\theta_1=\theta_2$ (i.e., geometrical optics). This maximum performance is also obtained by the proposed analytical optimization as well as the considered numerical one. In fact, it is also depicted that, when the RISs are optimized (analytically or numerically), the MI has approximately the same maximum value, irrespective of the lattice orientation.}
		\label{fig:MI_theta1_2RIS}
\end{figure}

The ergodic MI performance in nats per channel use (npcu) is illustrated in Fig$.$~\ref{fig:MI_AS_124RIS} as a function of the AS $\sigma$ in degrees for a TX-RX communication system operating at $2.5$ GHz and enabled by the deployment of $K=\{1,2,4\}$ RISs. We have assumed $\nt=8$ and $\nr=4$ for the number of TX and RX antenna elements, respectively, the SNR $\rho$ and the AS $\sigma$ were set to $10$ dB and $5^o$, respectively, and we have set $\theta_1=30^o$ and $\theta_2=70^o$ for the incoming and outgoing angles of the EM waves, respectively. Two cases have been simulated for this scenario where geometrical optics is not possible: the case where $\bPhi_k=\bI_{\ns}$ $\forall$$k\in\{1,2,4\}$ (i.e., unoptimized phase configurations for the RISs), and the case where $\bPhi_k$'s are optimized via both the proposal analytical approach and the numerical one. As observed in Fig$.$~\ref{fig:MI_AS_124RIS} for AS values up to about $20^o$, the optimization over each $\bPhi_k$ provides significant gains, irrespective of the size of each RIS. In fact, for ASs around $5^o-10^o$, the ergodic MI is higher than the one with negligible or zero antenna correlation (i.e., for large AS values). This surprising phenomenon is due to the significant ``beamforming'' gain resulting from the coherent reflection from the RISs along the directions of the non-negligible eigenvalues of the correlation matrices. As expected, increasing the number of RISs in the system increases the MI performance. It is also demonstrated that the performance of the proposed analytical optimization coincides with that of brute-force optimization.
\begin{figure}[!t]
\centering
\includegraphics[width=\columnwidth]{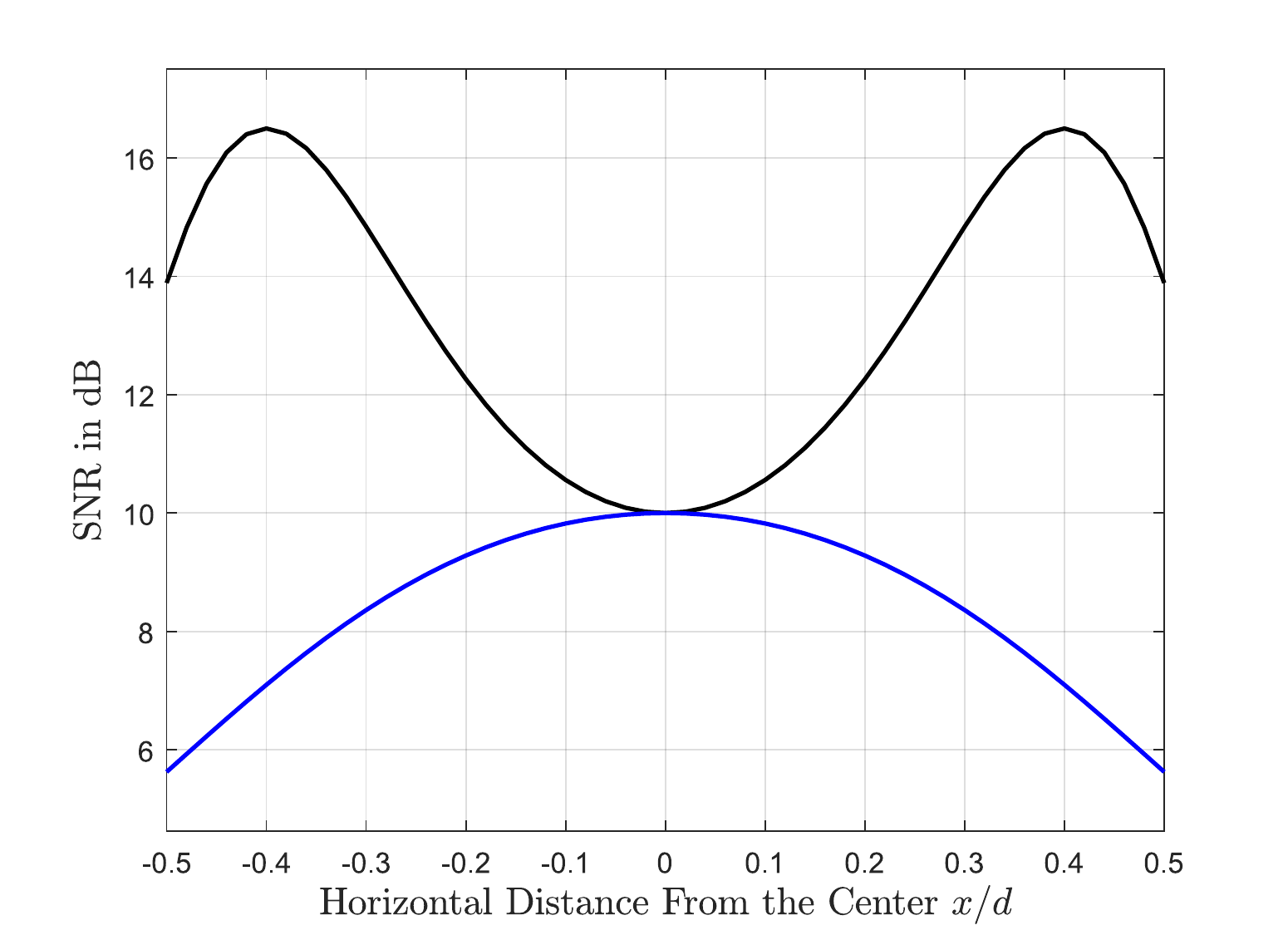}
\caption{The variation of the SNR in dB using \eqref{eq:gamma} versus the normalized horizontal distance $x/d$ of $K=2$ RISs from the center between the TX and RX locations. For every value of $x$, the two RISs are located  at $(x,\pm h)$ opposite from each other (see Fig. \ref{fig:system_model}). We have considered two values for $h$, namely  $h=0.7d$ (blue curves) and $h=0.3d$ (black curves).}
\label{fig:SNR}
\end{figure}

We next explore another important aspect of the optimization of the multiple RISs in the system, namely the effect of tilting the surfaces. As seen in the previous section and expression \eqref{eq:unit_rank_Sigma}, when the correlation matrices $\bS_{t,k}$'s and $\bS_{r,k}$'s are unit-rank, the degree to which optimization plays a role depends on whether the desired impinging and outgoing directions are those corresponding to geometrical optics. In Fig.~\ref{fig:MI_theta1_2RIS}, we plot the ergodic MI in the presence of $K=2$ RISs as a function of the incoming angle $\theta_1$, with respect to the vertical of each single RIS, for fixed desired angular difference between the incoming and outgoing mean directions (corresponding to the sum $\theta_1+\theta_2$ in the figure). The rest of the system parameters are the same with Fig$.$~\ref{fig:MI_AS_124RIS}. Note that, varying $\theta_1$ is equivalent to rotating (or tilting) each RIS around a horizontal axis. We see that, when $\theta_1=\theta_2$, which corresponds to geometrical optics, optimization gives no gains. The gains can be, however, significant for other orientations of the RISs. Interestingly, in both Figs.~\ref{fig:MI_AS_124RIS} and~\ref{fig:MI_theta1_2RIS}, brute-force optimization using \cite{Zhang_Capacity} produces identical results with the setting $[\mathbf{\Phi}_k]_{n,n}=\angle([\bu_{1,k}]_n[\bv_{1,k}^*]_n)$ for each $n$-th reflecting element with $n=1,2,\ldots,\ns$ of each $k$-th RIS. This corresponds to the phase difference between the elements of the eigenvectors corresponding to the maximum eigenvalues, as discussed in the previous section. 
\begin{figure}[!t]
\centering
\includegraphics[width=1.08\columnwidth]{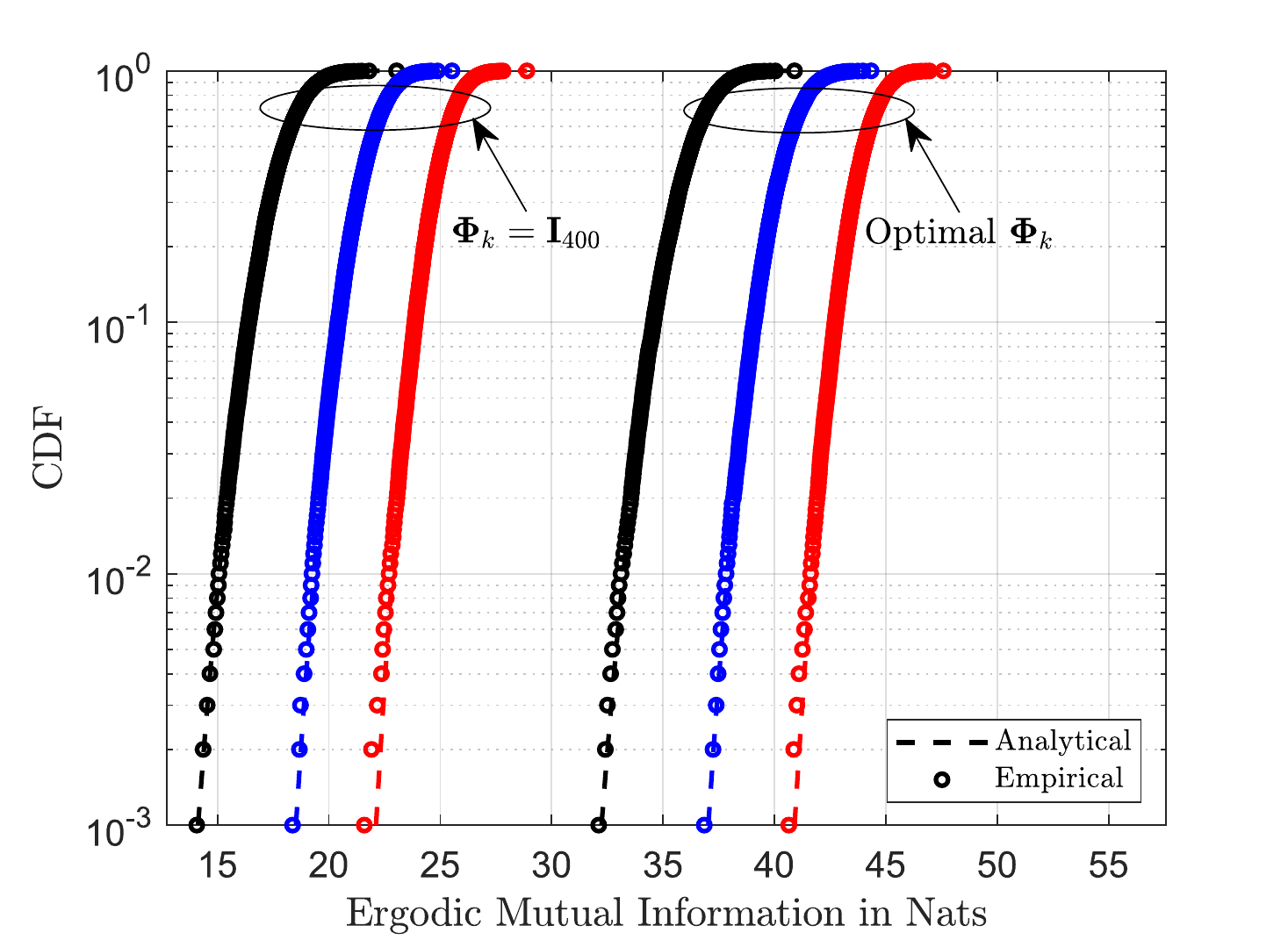}
\caption{The CDF of the ergodic MI in nats per channel use in the presence of $K=1$ (black), $K=2$ (blue) and $K=4$ (red) RISs, using the parameters' setting of Fig.~\ref{fig:MI_AS_124RIS}, i.e., for $\nt=8$, $\nr=4$, $\theta_1=30^o$, $\theta_2=70^o$, $\sigma=5^o$, and $\rho=10$ dB. The circled curves were generated with Monte Carlo simulations, while the dashed curves correspond to a Gaussian approximation using the mean and variance obtained analytically in Proposition~\ref{prop:ergMI}. The three curves on the left correspond to the unoptimized case with $\bPhi_k=\bI_{400}$ $\forall$$k$, while the three curves to the right have been obtained with optimized $\bPhi_k$'s, as per expression \eqref{eq:Phi_opt}. It is evident that the agreement between the theoretical and numerically generated distributions is sufficiently good. }
\label{fig:MI_CDFs}
\end{figure}

In addition to the above analysis, it is important to assess the relevance of the placements of the multiple RISs in the system. In this context, there are two factors that may play a role in the performance. First, the location of the RIS may affect the overall pathloss of the reflected signal. Interestingly, as discussed in Proposition \ref{prop:OptSNR}, the maximum SNR is not always at the center of the distance between the TX and RX antenna arrays. Depending on the horizontal distance $h$ (see Fig.~\ref{fig:system_model}), the maximum may be at the center (for $h>d/2$), or away from the center (when $h<d/2$). Furthermore, the location of each RIS, for a fixed orientation, may also play a role in its reflectivity, especially if the position of the RIS does not allow for geometrical optics reflection. In Fig.~\ref{fig:MI_x_2RIS}, we take both factors into account and plot the ergodic MI performance versus the horizontal distance $x/d$ from the TX-RX distance center, considering $K=2$ RISs located at $(x,\pm h)$ opposite from each other. We perform this analysis for two different values of $h$, one greater than $d/2$ and one less than $d/2$. 
Comparing the MI for the unoptimized phase configuration case (i.e., $\bPhi_k=\bI_{\ns}$ for $k=1$ and $2$) and the optimized one (where $\bPhi_k$'s are obtained using the closed-form expression \eqref{eq:Phi_opt}), we see that the RIS phase configuration optimization regains nearly the fully optimal value of the MI for any distance. This indicates that the optimization of the RISs makes the MI remain nearly constant as a function of distance, despite the variations of both the SNR and the waves' angles of arrival and departure at the RISs.

Finally, using the analytical results for the first two statistical moments of the MI metric in Section~\ref{sec:MI_Analysis}, and specifically \eqref{eq:Gaussian_approximation}, which capitalizing on Proposition~\ref{prop:ergMI}, approximates the MI distribution with a Gaussian one, we plot in Fig.~\ref{fig:MI_CDFs} the CDF of the MI, which is also compared with Monte-Carlo simulations for various values of $K$ for the numbers of the RISs. In this figure, we have used the same setting of parameters with Fig$.$~\ref{fig:MI_AS_124RIS}. It can be clearly seen that the agreement between the theoretical and numerically generated distributions is remarkable.

\section{Conclusions and Future Work}
\label{sec:conclusion}
In this paper, we have derived novel expressions for the mean and variance of the MI performance of multi-antenna wireless communication systems in the presence of multiple RISs, using random matrix theory and methods from statistical physics. While nominally valid in the limit of large numbers of TX and RX antennas and RIS elements, we have shown that treating the MI as a Gaussian variable, and neglecting higher cumulant moments which vanish in that limit, is quite accurate even for modest sized antenna arrays and RISs. The above results become particularly important when the channel is fast-fading, in which case it makes more sense to optimize the end-to-end communication link, and particularly the phases of the RISs' reflecting elements, based on statistical, rather than instantaneous, knowledge of the wireless channels.

In this case, we have shown that finite ASs can play a crucial role in the degree to which the RISs can be optimized. While for large AS values, optimization is unnecessary, for low AS values, which is reasonable for increasing carrier frequencies, which exhibit reduced multipath, significant capacity gains can be obtained with the optimization of the RISs. Correspondingly, phase configurations' optimization plays an important role when the required incoming and outgoing directions of the EM waves at the RIS are significantly different from the ones prescribed by geometrical optics. For example, we showcased that physical tilting of the RISs, to take advantage of geometrical optics, is unnecessary when {\it electrical} tilting, i.e., through phase optimization, is available. Moreover, we found that phase optimization of the RISs may have a far more profound effect to the ergodic MI improvement, as compared to their placement, due to huge gains in the number of channel modes participating in the propagation. 

Moving forward, it is important to analyze the potential gains and limitations on capacity, interference, as well as the complexity in the RISs' optimization in the presence of multiple TX-RX antenna arrays, sharing the wireless medium and the RISs, which lies at the heart of the success of RISs for 6G wireless communications. The model described in this paper can be readily applied to assess the impact of multiple RISs in range extension, propagation around obstacles, as well as the limitations of the use of RISs with multiple TX-RX pairs. 


\begin{appendices}
\section{Proof of Proposition \ref{prop:ergMI}}
\label{app:proof_ergMI}
To prove Proposition~\ref{prop:ergMI}, we will use methods from random matrix theory and the replica approach, a technique originally developed in
the context of statistical physics and successfully applied to several problems in wireless communications \cite{Tanaka2002_ReplicasInCDMAMUD, Moustakas2003_MIMO1, Guo2003_ReplicaAnalysisOfLargeCDMASystems, Muller2003_RandomMatrixMIMO_binary_mine, Taricco2008_MIMOCorrelatedCapacity}. Since the method has already been analyzed in the past, we will only provide highlights of the proof. We commence by defining the following scalar quantity:
\begin{equation}
    {\cal Z}\triangleq\det\left(\bI_\nr+\rho\bG_{\rm tot}\bQ\bG_{\rm tot}^\dagger\right)^{-1},
\end{equation}
using the corresponding letter ${\cal Z}$, which is used in Statistical Physics for the so-called partition function of the system. Then, the Moment Generating Function (MGF) of the MI can be obtained as $g(\nu)\triangleq\ex\left[{\cal Z}^\nu\right]$. From this expression, the following asymptotic formula for the normalized MI can be deduced:
\begin{equation}\label{eq:c=loggnu}
    C=-\lim_{\nt\to\infty} \nt^{-1}\left.\left( \log g(\nu)\right)'\right|_{\nu=0^+}.
\end{equation}
In addition, we can obtain the variance of the MI as follows:
\begin{equation}\label{eq:VarI=loggnu''}
    \text{Var}(I)=\lim_{\nt\to\infty} \left.\left( \log g(\nu)\right)''\right|_{\nu=0^+}.
\end{equation}
To proceed further, we will make a number of assumptions, which have been shown to hold in a related setting to this paper \cite{Guerra2002_ThermodynamicLimitSG, Talagrand2006_ParisiFormula}. In this context, it has been conjectured that they are valid\cite{Moustakas2003_MIMO1, Tanaka2002_ReplicasInCDMAMUD}. 
\begin{assumption}
The calculation of $g(\nu)$ for $\nu\in\ZZ^+$ can be analytically continued to real values of $\nu\in(0,1)$.
\end{assumption}
\begin{remark*}\rm
This property will allow us to obtain the behavior close to $\nu=0^+$ by evaluating the expression for integer values of the replica index $\nu$.
\end{remark*}
We will now start with the evaluation of $g(\nu)$ for $\nu\in\ZZ^+$. It is straightforward to show the following expression (see \cite{Moustakas2007_MIMO1} for details):
\begin{align}
    &{\cal Z}^\nu = \ex\left[
    e^{\frac{1}{2}\Tr{\bY^\dagger\bG_{d}^\dagger  \bZ-\bZ^\dagger \bG_{d}\bQ\bY }}
\times\right. 
\\ \nonumber
&\left.\!
         e^{\frac{1}{2}\!\!\sum\limits_{k=1}^K\!\!\Tr{\bV_k^\dagger\bG_{t,k}\bY-\bY^\dagger \bQ\bG_{t,k}^\dagger\bW_k-\bZ^\dagger\bG_{r,k}\bPhi_k\bV_k-\bW_k^\dagger\bPhi^\dagger_k\bG_{r,k}^\dagger\bZ}}
        \right]\!\!,
\end{align}
where the expectation is over the zero-mean complex Gaussian matrices $\bZ$, $\bY$, $\bV_k$, and $\bW_k$ ($k=1,2,\ldots,K$) of dimensions $\nr\times \nu$, $\nt\times\nu$, $\ns\times\nu$ and $\ns\times\nu$, respectively, each with variance equal to $2$. In the above expression, the channel matrices appear in the exponent of the exponential in a linear fashion, and hence, they may be integrated out resulting to:
\begin{align}
\label{eq:gnu1}
&g(\nu) =  \ex\left[ e^{ -\frac{1}{4\nt}\Tr{\bY^\dagger\bQ\bT_d\bY\bZ^\dagger\bR_d\bZ} } 
\right.
\\ \nonumber
&\left.\!\times 
e^{\frac{1}{4\nt} \sum_{k=1}^K\Tr{\bZ^\dagger\bR_k\bZ\bW_k^\dagger\bPhi_k^\dagger\bS_{rk}\bPhi_k\bV_k - \bY^\dagger\bQ\bT_k\bY\bV_k^\dagger\bS_{tk}\bW_k}} \right]\!\!.
\end{align}
The integration over the Gaussian channel matrices has resulted into terms in the exponent, which are quartic in the Gaussian random variables. To overcome this difficulty we employ the following identity, which decomposes them into quadratic terms via the Fourier representation of the Dirac $\delta$-function.
\begin{identity}
\label{id:hub_strat}%
If $\bA, \bB\in\CC^{\nu\times\nu}$, the following identity holds \cite{Taricco2008_MIMOCorrelatedCapacity}:
\begin{align}
\label{eq:hub_strat_identity}%
&e^{-\frac{1}{\nt}\Tr{\bA\bB}}  =
\\ \nonumber
& \lim_{\epsilon\to 0^+}\int e^{ \Tr{\nt\left(-\epsilon{\mathbfcal{T}  \mathbfcal{T}}^T+\epsilon{\mathbfcal{R} \mathbfcal{R}}^T+{\mathbfcal{R} \mathbfcal{T}}\right)-  \bA {\mathbfcal  T} - {\mathbfcal R}
\bB }}d\mu({\mathbfcal  T}, {\mathbfcal
R}),
\end{align}
where the integration metric $d\mu({\mathbfcal T, \bRcal})$ is given by
\begin{equation}
d\mu({\mathbfcal{T}, \mathbfcal{R}}) = \prod_{\alpha,\beta=1}^\nu \nt 
\frac{d[{\mathbfcal T}]_{\alpha\beta} d[{\mathbfcal R}]_{\beta\alpha}}{2\pi i}    
\end{equation}
and the integration of the ${\mathbfcal T}$-matrix elements is over the real axis, while the integration of the ${\mathbfcal R}$ is over the imaginary axis.
\end{identity}

Following a standard procedure \cite{Moustakas2003_MIMO1}, we introduce the matrices: \textit{i}) ${\mathbfcal T}_d$ and ${\mathbfcal R}_d$ to decompose the term in the first line of \eqref{eq:gnu1} with $\bA=\frac{1}{2\nt}\bZ^\dagger\bR_d\bZ$ and $\bB=\frac{1}{2}\bY^\dagger\bQ\bT_d\bY$; \textit{ii}) the matrices ${\mathbfcal T}_{1k}$ and ${\mathbfcal R}_{1k}$, with $k=1,2,\ldots,K$, to decompose the first term in the second line of \eqref{eq:gnu1} with $\bA=-\frac{1}{2}\bW_k^\dagger\bPhi_k^\dagger\bS_{rk}\bPhi_k\bV_k$ and $\bB=\frac{1}{2\nt}\bZ^\dagger\bR_k\bZ$; and finally \textit{iii}) the matrices ${\mathbfcal T}_{2k}$, ${\mathbfcal R}_{2k}$, with $k=1,2,\ldots,K$, to decompose the second term in the second line of \eqref{eq:gnu1} with $\bA=\frac{1}{2\nt}\bY^\dagger\bQ\bT_k\bY$ and $\bB=\frac{1}{2}\bV_k^\dagger\bS_{tk}\bV_k$. 
This allows us to integrate over the complex matrices $\bZ$, $\bY$, $\bV_k$, and $\bW_k$. Hence, \eqref{eq:gnu1} can be compactly re-expressed as:
\begin{align}
\label{eq:gnu2} %
g(\nu)=\int e^{-{\cal S }}d\mu(\{\bTcal, \bRcal\}),
\end{align}
with $d\mu(\{\bTcal, \bRcal\})$ indicating an integration over all three versions of the $\mathbfcal{T}$ and $\bRcal$ matrices introduced above, and where the exponent ${\cal S}$ takes the following form:
\begin{align}
 &{\cal S} \triangleq \log\det \left(\bI_\nr \outer \bI_{\nu} + \left(\bR_d \outer \bRcal_d+
  \sum_{k=1}^K \bR_k \outer \bRcal_{1k} \right)\right)
\nonumber \\
&+\log\det \left(\bI_\nt \outer \bI_{\nu} + \rho\left(\bQ\bT_d \outer \bTcal_d+
  \sum_{k=1}^K \bQ\bT_k \outer \bTcal_{2k} \right)\right)
\nonumber \\
 &+\sum_{k=1}^K \log\det \left(\bI_\nu \outer \bI_{\ns} +  
\bTcal_{1k}\bRcal_{2k}\outer \bPhi_k\bS_{rk}\bPhi_k\bS_{tk}  \right)
\nonumber \\%
 &- \nt\Tr{\bTcal_d \bRcal_d+ \sum_{k=1}^K \left(\bTcal_{1k} \bRcal_{1k} + \bTcal_{2k} \bRcal_{2k}\right) },
\label{eq:S_full}
\end{align}
where the notation $\outer$ denotes the direct product of matrices. Since the integral in \eqref{eq:gnu2} cannot be performed exactly, we will evaluate it asymptotically for large numbers of antenna (TX/RX) and metamaterial/reflecting (RISs) elements.
\begin{assumption}
The above analytic continuation of $g(\nu)$ to real values of $\nu$ in \eqref{eq:c=loggnu} can be interchanged with the limit $\nt\to\infty$.
\end{assumption}
To obtain the asymptotic evaluation of \eqref{eq:gnu2}, we will deform the contours of the integrals of the elements of $\{\mathbfcal{T}, \mathbfcal{ R}\}$ to pass through a saddle point of ${\cal S}$. More details can be found in \cite{Moustakas2003_MIMO1, Bender_Orszag_book}. To proceed, we need to specify the structure of the saddle-point and the symmetry of the values of the matrices there. Since the $\nu$ replicas are introduced a-priori equivalent with each other, it is natural to make the following assumption: 
\begin{assumption}
At the saddle-point, the matrices $\{\mathbfcal{T}, \mathbfcal{ R}\}$ appearing in \eqref{eq:S_full} are rotationally invariant under continuous replica rotations and are, thus, proportional to $\bI_\nu$.
\end{assumption}
\begin{remark*}
\rm This technical assumption, which seems intuitively obvious, very often becomes invalid in certain systems, in which this symmetric solution becomes unstable, leading to the so-called replica-symmetry breaking phenomenon \cite{ParisiBook}. Nevertheless, in this case, due to the continuous symmetry present, the symmetry can be shown to be always stable \cite{Moustakas2007_MIMO1}, giving extra credence to the result.
\end{remark*}
Given the aforepresented assumption, we have:
\begin{align}
    \left.{\mathbfcal T}_d\right|_{saddle point}=t_d\bI_\nu &\mbox{,}\,\,\,\,\,\,\,\,\,\, \left.{\mathbfcal R}_d\right|_{saddle point}=r_d\bI_\nu, 
    \\ \nonumber
    \left.{\mathbfcal T}_{ak}\right|_{saddle point}=t_{ak}\bI_\nu &\mbox{,}\,\,\,\,\,\,\,\,\,\, \left.{\mathbfcal R}_{ak}\right|_{saddle point}=r_{ak}\bI_\nu, 
\end{align}
where $a=1,2$ and $k=1,2,\ldots,K$. To obtain the values of $r_d$, $t_d$, $r_{ak}$, and $t_{ak}$, we need to solve the saddle-point equations, which can be derived by demanding that ${\cal S}$ is stationary with respect to variations of the matrices $\{{\mathbfcal T, \bRcal}\}$ (see \cite{Bender_Orszag_book} for details). This produces the fixed-point equations in \eqref{eq:fp_eqs}, which can be shown to have unique solutions \cite{Taricco2008_MIMOCorrelatedCapacity}. Hence, $g(\nu)$ can be written as follows:
\begin{equation}
    g(\nu)=e^{-{\cal S}_0} \int e^{-({\cal S}-{\cal S}_0)} d\mu(\{{\mathbfcal{T}, \mathbfcal{R}}\}),
\end{equation}
where ${\cal S}_0\triangleq\nu \nt C$ is the saddle-point value of ${\cal S}$, and $C$ is given in \eqref{eq:S0}. We will now show how to obtain the variance from the asymptotic evaluation of the above integral  \cite{Moustakas2003_MIMO1, Moustakas2007_MIMO1}. Since ${\cal S}=O(\nt)$, for large $\nt$, the above integral is dominated by the value of the integrand at its saddle point. Hence, we may expand ${\cal S}$ around its saddle-point value and integrate the leading quadratic term, treating the remainder as perturbation, which can be shown to be $O(\nt^{-1})$. Therefore, we have that:
\begin{eqnarray}\label{eq:S_expansion}
{\cal S}={\cal S}_0+\nt {\cal S}_2+ \nt \sum_{\ell=3}^\infty {\cal S}_\ell,
\end{eqnarray}
where  ${\cal S}_2$ is given by:
\begin{align}\nonumber 
    &{\cal S}_2 = -\sum_k\Tr{\delta{\mathbfcal T}_{1k} \delta{\mathbfcal R}_{1k}\!+\!\delta{\mathbfcal T}_{2k} \delta{\mathbfcal R}_{2k}\!+\!\delta{\mathbfcal T}_{d} \delta{\mathbfcal R}_{d}}
    \\ \nonumber 
    +&\sum_{k=1}^K[{\bf M}_{12}]_{kk} \Tr{\delta{\mathbfcal T}_{1k} \delta{\mathbfcal R}_{2k}}      
    \\ \nonumber   
    +&\frac{1}{2}\! 
    \sum_{k,k'=1}^K \left([{\bf M}_{1r}]_{kk'}
    \Tr{ \delta\mathbfcal{R}_{1k}\delta\mathbfcal{R}_{1k'}} \!+\! [{\bf M}_{1t}]_{kk'} 
    \Tr{ \delta\mathbfcal{T}_{1k}\delta\mathbfcal{T}_{1k'}}\right)
    \\ \nonumber
    +&\frac{1}{2}\!
    \sum_{k,k'=1}^K \left([{\bf M}_{2r}]_{kk'}
    \Tr{ \delta\mathbfcal{R}_{2k}\delta\mathbfcal{R}_{2k'}} \!+\! [{\bf M}_{2t}]_{kk'} 
    \Tr{ \delta\mathbfcal{T}_{2k}\delta\mathbfcal{T}_{2k'}} \right)
    \\ \nonumber
    +&\sum_{k=1}^K \left([{\boldsymbol \mu}_{d1r}]_{k}
    \Tr{ \delta\mathbfcal{R}_{1k}\delta\mathbfcal{R}_{d}} 
    \!+\! [{\boldsymbol \mu}_{d2t}]_{k}
    \Tr{ \delta\mathbfcal{T}_{2k}\delta\mathbfcal{T}_{d}} \right)
    \\
    +&\frac{1}{2} 
    \left( \varrho_{dr} 
    \Tr{ \delta\mathbfcal{R}_{d}^2} 
    \!+\! \varrho_{dt} 
    \Tr{\delta\mathbfcal{T}_{d}^2} \right).
\end{align}
In the last expression, $\delta {\mathbfcal R}_{1k} = {\mathbfcal R}_{1k}-\left.{\mathbfcal R}_{1k}\right|_{saddlepoint}$, and the similar bold-faced matrix notations, represent $\nu\times\nu$ matrices. It is noted that, in \eqref{eq:S_expansion}, all terms and ${\cal S}_\ell$ for $\ell\geq 3$ include expansion terms, which are polynomial in $\delta {\mathbfcal R}_{1k}$, $\delta {\mathbfcal R}_{1k'}$, $\delta {\mathbfcal R}_{2k}$, and $\delta {\mathbfcal R}_{2k'}$ of degree $\ell$. Furthermore, the $K\times K$ matrices ${\bf M}_{1r}$, ${\bf M}_{2r}$, ${\bf M}_{1t}$, ${\bf M}_{2t}$, and ${\bf M}_{12}$, and the $K$-dimensional vectors ${\boldsymbol \mu}_{1dr}$ and ${\boldsymbol \mu}_{2dt}$ have elements defined as follows:
\begin{align}
    [{\bf M}_{1r}]_{kk'} &\triangleq -\frac{1}{\nt} \Tr{ \left({\bf I}_{\nr}+\tilde{\bf R}\right)^{-1}  {\bf R}_k\left({\bf I}_{\nr}+\tilde{\bf R}\right)^{-1} {\bf R}_{k'}}, 
    \nonumber \\ 
    [{\bf M}_{2r}]_{kk'} &\triangleq -\delta_{kk'}\frac{\gamma_k^2t_{1k}^2}{\nt} \Tr{ \left({\bf I}_{\ns}+\gamma_k t_{1k}r_{2k}{\bf \Sigma}_k\right)^{-2}  {\bf \Sigma}_k^2 },
    \\ \nonumber
    [{\bf M}_{1t}]_{kk'} &\triangleq -\delta_{kk'}\frac{\gamma_k^2r_{2k}^2}{\nt} \Tr{ \left({\bf I}_{\ns}+\gamma_k t_{1k}r_{2k}{\bf \Sigma}_k\right)^{-2}  {\bf \Sigma}_k^2 },
    \\ \nonumber
    [{\bf M}_{2t}]_{kk'} &\triangleq - \frac{\rho^2}{\nt} \Tr{ \left({\bf I}_{\nt}+\tilde{\bf T}\right)^{-1}  {\bf QT}_k\left({\bf I}_{\nt}+\tilde{\bf T}\right)^{-1} {\bf QT}_{k'}},
    \\ \nonumber
    [{\bf M}_{12}]_{kk'} &\triangleq \delta_{kk'}\frac{\gamma_k}{\nt} \Tr{ \left({\bf I}_{\ns}+\gamma_k t_{1k}r_{2k}{\bf \Sigma}_k\right)^{-2}  {\bf \Sigma}_k },
    \\ \nonumber
    [{\boldsymbol \mu}_{1dr}]_{k} &\triangleq -\frac{1}{\nt} \Tr{ \left({\bf I}_{\nr}+\tilde{\bf R}\right)^{-1}  {\bf R}_k\left({\bf I}_{\nr}+\tilde{\bf R}\right)^{-1} {\bf R}_{d}}, 
    \\ \nonumber
    [{\boldsymbol \mu}_{2dt}]_{k}&\triangleq - \frac{\rho^2}{\nt} \Tr{ \left({\bf I}_{\nt}+\tilde{\bf T}\right)^{-1}  {\bf QT}_k\left({\bf I}_{\nt}+\tilde{\bf T}\right)^{-1} {\bf QT}_{d}},
\end{align}
and the scalars $\varrho_{dr}$ and $\varrho_{dt}$ are given by:
\begin{align}
    \varrho_{dr} &\triangleq -\frac{1}{\nt} \Tr{ \left(\left({\bf I}_{\nr}+\tilde{\bf R}\right)^{-1}  {\bf R}_d\right)^2}, 
    \\ \nonumber
    \varrho_{dt}&\triangleq  -\frac{\rho^2}{\nt} \Tr{ \left(\left({\bf I}_{\nt}+\tilde{\bf T}\right)^{-1}  {\bf QT}_d\right)^{2} }.
\end{align}

We next integrate over all the elements of the matrices $\delta\mathbfcal{T}$ and $\delta\mathbfcal{R}$, deforming the path of each variable so as to approach the saddle point from the stationary phase direction. More details about this procedure can be found in \cite{Moustakas2007_MIMO1}. We thus obtain the following expression for the MGF of the MI:
\begin{equation}
    g(\nu)=e^{-{\cal S}_0} 
    \det\left({\bf \Lambda}\right)^{-\frac{\nu^2}{2}},
\end{equation}
where the $(4K+2)$-dimensional matrix ${\bf \Lambda}$ is given by
\begin{align}\label{eq:Vmat_def}
    {\bf \Lambda} = \left[\begin{array}{cccccc} 
    \varrho_{dt} & 0 & {\boldsymbol \mu}_{2dt} & 
    -1 & 0 & 0
    \\
    0 & {\bf M}_{1t} & 0 & 
    0 & -\bI_K & {\bf M}_{12}
    \\
    {\boldsymbol \mu}_{2dt} & 0 & {\bf M}_{2t} & 
    0 & 0 & -\bI_K
    \\
    -1 & 0 & 0 & 
    \varrho_{dr} & {\boldsymbol \mu}_{1dr} & 0
    \\
    0 & -\bI_K & 0 & 
    {\boldsymbol \mu}_{1dr} & {\bf M}_{1r} & 0
    \\
    0 & {\bf M}_{12} & -\bI_K & 
    0 & 0 & {\bf M}_{2r}
    \end{array}\right]
\end{align}
Combining the above equations with \eqref{eq:VarI=loggnu''}, yields \eqref{eq:Var(I)}.

If we follow the approach of \cite{Moustakas2003_MIMO1, Moustakas2007_MIMO1} and evaluate the terms ${\cal S}_\ell$ for $\ell\geq 3$ in \eqref{eq:S_expansion} in a perturbative manner around the above saddle point, we may obtain the skewness and all other higher-order moments of the MI metric. Simple power counting over $\nt$ can then show that all MI's higher moments vanish in the large $\nt$ limit, thus proving that, in that limit, the distribution of the MI is asymptotically Gaussian.

\end{appendices}

\bibliographystyle{IEEEtran}
\bibliography{IEEEfull,wireless,references}

\end{document}